\documentclass[11pt]{article}
\usepackage{mathrsfs}
\usepackage{amsmath, alltt}
\usepackage{rotating}
\usepackage{graphicx}
\usepackage{subfigure}
\usepackage[onehalfspacing]{setspace}
\usepackage[hmargin={1.0in, 1.0in}, vmargin={1.0in, 1.0in}]{geometry}
\usepackage{amsfonts}
\usepackage{amssymb}
\usepackage{natbib}
\usepackage{amsmath}
\usepackage{bm}
\usepackage{latexsym}
\usepackage{amssymb,bm}
\usepackage[section]{placeins}
\usepackage{longtable}
\usepackage{lscape}
\usepackage{scalefnt}
\usepackage{setspace}
\usepackage{color}
\usepackage{mathtools}
\usepackage{enumitem}

\newcommand{\blind}{1}

\begin{document}
	\setlength{\abovedisplayskip}{6pt}
	\setlength{\belowdisplayskip}{6pt}
	\setlength{\abovedisplayshortskip}{6pt}
	\setlength{\belowdisplayshortskip}{6pt}	
\if1\blind
{
	\title{Regression for Copula-linked Compound Distributions with Applications in Modeling Aggregate Insurance Claims}	
	\author{
		Peng Shi \\
		Department of Risk and Insurance\\
		Wisconsin School of Business\\
		University of Wisconsin-Madison\\
		Email: pshi@bus.wisc.edu\\
		\and
		Zifeng Zhao \\
		Department of Information Technology, Analytics, and Operations \\
		Mendoza College of Business\\
		University of Notre Dame\\
		Email: zzhao2@nd.edu
	}
	\date{}	
	\maketitle
} \fi

\if0\blind
{
	\title{Regression for Copula-linked Compound Distributions with Applications in Modeling Aggregate Insurance Claims}	
	\author{}
	\date{}
	\maketitle
} \fi

\bigskip

\begin{abstract}

In actuarial research, a task of particular interest and importance is to predict the loss cost for individual risks so that informative decisions are made in various insurance operations such as underwriting, ratemaking, and capital management. The loss cost is typically viewed to follow a compound distribution where the summation of the severity variables is stopped by the frequency variable. A challenging issue in modeling such outcome is to accommodate the potential dependence between the number of claims and the size of each individual claim. In this article, we introduce a novel regression framework for compound distributions that uses a copula to accommodate the association between the frequency and the severity variables, and thus allows for arbitrary dependence between the two components. We further show that the new model is very flexible and is easily modified to account for incomplete data due to censoring or truncation. The flexibility of the proposed model is illustrated using both simulated and real data sets. In the analysis of granular claims data from property insurance, we find substantive negative relationship between the number and the size of insurance claims. In addition, we demonstrate that ignoring the frequency-severity association could lead to biased decision-making in insurance operations.

%

\end{abstract}

\noindent
{\it Keywords:} Aggregate insurance claims, Compound distributions, Copula regression, Incomplete data, Two-part models

\newpage
\section{Introduction} \label{sec:intro}

In actuarial research on nonlife insurance, a task of particular interest and importance is to predict the loss cost for individual risks in an insurer's book of business. Interpretation and prediction of loss cost of individual policyholders deepens the insurer's understanding of the risk profile of the entire portfolio, which further leads to better-informed decisions in various insurance operations such as underwriting, ratemaking, and capital management.

The loss cost of a policyholder is jointly determined by the number of claims and the amount of each claim during the contract period. As a result, researchers and practitioners typically view the loss cost outcome to follow a compound or generalized distribution (see \cite{KarlisEvdokia2005} and \cite{JohnsonKotz2005}). Specifically, the loss cost per policy year, denoted by $S$, can be represented as:
\begin{align} \label{equ:crm}
S = Y_1+\cdots + Y_N,
\end{align}
where $N$ is a counting random variable and represents the number of claims, and $Y_j$ ($j=1,\ldots,N$) is a non-negative continuous random variable and represents the size of the $j$th claim. The sequence of $Y_1, Y_2, \cdots$ is further assumed to be independently and identically distributed. Compound distributions have been extensively used in the actuarial science literature for modeling aggregate losses in an insurance system (see, for example, \cite{Klugman2012}, \cite{Lin2014}, and \cite{Albrecher2017}). In insurance applications, $N$ and $\{Y_j\}$ are referred to as the ``frequency'' and ``severity'' components respectively.



In this article, we focus on the regression method for compound distributions when both $N$ and $(Y_1,\ldots,Y_N)$ are observed. A challenging issue in modeling such outcomes in the regression setting is to accommodate the potential dependence between the number of claims and the size of each individual claim. The goal of this work is to introduce a simple yet flexible regression framework to allow for arbitrary dependence between the frequency and severity distributions.


The current regression approach to studying the aggregate loss $S$ relies on the independence assumption between $N$ and each $Y_j$. Under such independence assumption, one develops regression models for the number and size of claims separately, which is known as the frequency-severity or two-part model. See \cite{Frees2014} for discussions on various types of two-part models. As a special case, when the frequency is a Poisson variable and the severity is a gamma variable, the loss cost is known to follow a Tweedie distribution (\cite{Tweedie1984}). \cite{JorgensenSouza1994} and \cite{SmythJorgensen2002} have explored fitting the Tweedie's compound Poisson model to the loss cost data in property insurance.

In addition to actuarial science and insurance, regression models based on compound distributions have been used in many other disciplines as well. In health economics, the two-part model was used to study an individual's total number of doctor visits resulting from multiple spells of illness in a given period (see, for instance, \cite{SilvaWindmeijer2001}). In marketing, \cite{Tellis1988} employed a special case of the frequency-severity model to study the effect of repetitive advertising on consumer purchasing choices; \cite{AribargPietersWedel2010} studied consumer advertisement recognition where an individual's attention is formulated as a compound model determined by eye fixation frequency and gaze duration. In operational risk, the compound distribution for aggregate losses is the foundation for the determination of the operational risk capital required by the Basel
capital framework for banks (\cite{Panjer2006} and \cite{Shevchenko2010}). In psychology, \cite{SmithsonShou2014} pointed out the applications of this type of model in different areas of psychology such as perception and decision making, where a psychological process is thought to be serially summed from observable component process outputs.

The two-part models in different scientific fields described above employ some common key assumptions, including:
\begin{enumerate}[nolistsep]
\item[(1)] The distribution of $N$ does not depend on the values of $Y_j$ for $j=1,\ldots,N$;
\item[(2)] Conditional on $N=n>0$, $Y_1,\ldots,Y_n$ are independently distributed random variables;
\item[(3)] Conditional on $N=n>0$, the common distribution of $Y_1,\ldots,Y_n$ does not depend on $n$.
\end{enumerate}
The (conditional) independence assumption between $N$ and $Y_j$ certainly leads to tractable statistical inference because it allows one to build regression models separately for the frequency and severity components. However, if $N$ and $Y_j$ are correlated, ignoring the association between them will lead to serious biases in the inference. First, the regression coefficients in the severity regression model will be inconsistent estimates of the marginal effect of explanatory variables. Second, there is a persistent error in the prediction for the severity given the frequency. Third, the misspecification will introduce bias in the inference for the compound distribution.

Motivated by the above observations, we introduce a novel copula-linked compound distribution and the associated two-part regression framework that allow for arbitrary dependence between the frequency and severity components. Specifically, we employ a parametric copula to construct the joint distribution of frequency and severity variables, thus relax the independence assumption in standard methods. We show that the resulting copula regression framework is able to nest several commonly used approaches as special cases, including the hurdle model, the selection model, and the frequency-severity model, among others. Furthermore, we extend the basic model to accommodate the case of incomplete data due to censoring or truncation. Because of the parametric nature, likelihood-based approaches are proposed for estimation, inference, and diagnostics.

The flexibility of the proposed model is illustrated using both simulated and real data sets. In the numerical experiments, we showcase the impact of ignoring the frequency and severity dependence on the resulting compound distribution. In the empirical study, we apply the proposed method to granular claims data in property insurance. Our analysis detects substantive negative dependency between the number and the size of insurance claims. In addition, we demonstrate the importance of such dependency in some key insurance functions, including underwriting and ratemaking, loss reserving, and capital management. The results suggest that ignorance of frequency-severity dependence could lead to biased decision-making in insurance operations.

To the best of our knowledge, this work is among the first efforts to explicitly incorporate the dependence between the frequency and severity variables of a compound distribution in a regression setting. Recent literature has made some development in this direction, for example, see \cite{Czado2012mixed}, \cite{kramer2013total}, and \cite{Garrido2016} among others. The fundamental difference between our work and existing studies is that the aforementioned studies examined the relation between the frequency $N$ and the average severity $\overline{Y}=\sum_{j=1}^{N}Y_j/N$, while the proposed method directly looks into the relation between the frequency $N$ and the individual severity $Y_j$. Alternative mechanisms for introducing dependence between the frequency and individual severity variables include the correlated random-effect framework as in \cite{OlsenSchafer2001} and the conditional approach as in \cite{frees2011predicting}. The difficulty with both methods compared to the proposed copula approach is that it is not straightforward to handle incomplete data which is not unusual in insurance applications because of various coverage modifications.


Given that our work fits in the broader literature on multivariate modeling in insurance, it is worth discussing their differences and connections. The current literature on dependence modeling of insurance claims focuses on the joint modeling of multiple outcomes of loss cost that could arise from multiple lines of business (see \cite{FreesGeeLu2016}), multiple coverage in a single business line (see \cite{ShiFengBoucher2016}), or multiple peril types covered by a policy (see \cite{ShiLu2018}). In this line of studies, each loss cost outcome is formulated using either a Tweedie model or a two-part model. Both can be viewed in the framework of the compound distribution (\ref{equ:crm}) where the $N$ and each $Y_j$ are assumed to be independent with each other. Apart from the association among multiple loss cost outcomes, this work examines a single loss cost outcome, and the focus is on the dependence between the frequency and severity components in the compound model.

The rest of the paper is structured as follows. Section 2 introduces the dependent frequency-severity regression model for the compound distribution and discusses its extension for incomplete data due to censoring and truncation. The likelihood-based methods for estimation, inference, and diagnostics are further discussed. Section 3 provides numerical experiments to show the impact of ignoring the frequency-severity dependence under various settings. Section 4 applies the proposed approach to the loss cost data in property insurance and shows the importance of the frequency-severity dependence in insurance operations. Section 5 concludes the article. The supplementary materials contain additional technical examples, numerical studies, and detailed data analysis.

\section{Copula-linked Compound Regression} \label{sec:copulamodel}
\subsection{Basic Model}

In the basic setup, we assume that complete information on $(N,Y_1,\ldots,Y_N)$ is observed for each subject, where $N$ is a count variable, and $\{Y_j\}$ are continuous variables. 
For simplicity, we suppress the subject index in the following presentation. The joint distribution of $(N,Y_1,\ldots,Y_N)$ is built upon the assumption that $(Y_1,\ldots,Y_n)$ are conditionally i.i.d. given $N=n$ as opposed to the unconditional i.i.d. assumption in the standard compound distribution. There are several implications of this assumption. First, conditional independence of $(Y_1,\ldots,Y_n)$ given $N=n$ introduces correlation among $Y_j$, which departs from the i.i.d. assumption in the standard model; Second, identical distribution of $(Y_1,\ldots,Y_n)$ given $N=n$ implies identical marginal distribution of $Y_j$, which is consistent with the i.i.d. assumption in the standard model; Third, the bivariate distribution of $(N,Y_j)$ are identical given $N=n$, which nests the independent case in the standard model.

To facilitate presentation, we denote $Y$ as the variable associated with the common distribution of the sequence $\{Y_j\}$. Note that $Y$ is only defined in the sense of a distribution, not in the sense of a random variable. Under the conditional independence assumption, 
the associated pmf/pdf function is:
\begin{align}
f_{N,\bm{Y}}(n,y_1,\ldots,y_n) &= \left\{
                                    \begin{array}{ll}
                                      {\rm Pr}(N=0) & n=0 \\
                                      \frac{\partial}{\partial y_1\cdots \partial y_n}{\rm Pr}(N= n, Y_1\leq y_1,\cdots, Y_n \leq y_n)  & n>0
                                    \end{array}
                                  \right. \nonumber\\
&= \left[f_N(0)\right]^{I(n=0)}\left[f_N(n)\times f_{\bm{Y}|N}(y_1,\ldots,y_n|n)\right]^{I(n>0)} \nonumber\\
 &= f_N(n)\times \left[\prod_{j=1}^{n}f_{Y|N}(y_j|n)\right]^{I(n>0)},   \label{equ:coprisk}
\end{align}
where $I(\cdot)$ is an indictor function. 

The central component to define (\ref{equ:coprisk}) is the joint distribution of $N$ and $Y$. To allow for flexible dependence between $N$ and $Y$, \textcolor{black}{we take a parametric approach and employ a bivariate parametric} copula to construct their joint distribution. Refer to \cite{Nelsen2006} and \cite{Joe2015} for an introduction to dependence modeling with copulas.
According to the Sklar's theorem, the joint distribution of $N$ and $Y$ can be expressed in terms of a bivariate copula $C$:
\begin{align} \label{equ:cdf}
F_{N,Y}(n,y) = {\rm Pr}(N\leq n, Y\leq y) = C(F_N(n),F_Y(y)).
\end{align}
Denote $h(u,v)=\frac{\partial}{\partial v}C(u,v)$, it follows that
\begin{align} \label{equ:pdf}
f_{N,Y}(n,y) &= \frac{\partial}{\partial y}{\rm Pr}(N= n, Y\leq y) \nonumber \\
& = f_Y(y)[h(F_N(n),F_Y(y))-h(F_N(n-1),F_Y(y))].
\end{align}
From above, one finds the conditional distribution of $Y$ given $N$ as:
\begin{align}
F_{Y|N}(y|n) &= {\rm Pr}(Y\leq y|N=n) \nonumber \\
&=\dfrac{1}{f_N(n)}[C(F_N(n),F_Y(y))-C(F_N(n-1),F_Y(y))],  \\
f_{Y|N}(y|n) &= \frac{\partial}{\partial y}{\rm Pr}(Y\leq y|N=n)  \nonumber\\
& = \dfrac{f_Y(y)}{f_N(n)}[h(F_N(n),F_Y(y))-h(F_N(n-1),F_Y(y))].
\end{align}

In a regression context, one wants to incorporate exogenous explanatory variables to account for observed heterogeneity in both $N$ and $Y_j$. Thus, the marginal models for both $N$ and $Y_j$ are defined conditional on covariates. For example, in generalized linear models, one could specify $g^f({\rm E}(N_i|\bm{x}_i))=\bm{x}'_i\bm{\beta}^f$ and $g^s({\rm E}(Y_{ij}|\bm{x}_i))=\bm{x}'_i\bm{\beta}^s$, where $i$ is the subject index, $\bm{x}_i$ is the vector of covariates, $\bm\beta$ is the regression coefficients, and $g$ denotes the link function. Superscripts $f$ and $s$ indicate the frequency and severity components respectively.

As a special case, when the copula in (\ref{equ:cdf}) is an independence copula, i.e., $N$ and each $Y_j$ are independent, model (\ref{equ:coprisk}) reduces to:
\begin{align}
f_{N,\bm{Y}}(n,y_1,\ldots,y_n) = f_N(n) \times \left[\prod_{j=1}^{n} f_Y(y_j)\right]^{I(n>0)}   \label{equ:indrisk},
\end{align}
where the marginal models of $N$ and $Y$ are totally separable. Since (\ref{equ:coprisk}) nests (\ref{equ:indrisk}) as a special case, the usual goodness-of-fit statistics such as the likelihood ratio test could be used to test whether the independence assumption between $N$ and $Y_j$ is supported by the data.


It is worth stressing several observations in model (\ref{equ:coprisk}).  
First, the independence assumption of $Y_j$ given $N$ implies a specific dependence among the sequence $\{Y_j\}$. As pointed out by \cite{liu2017collective}, other types of dependence might exists between $N$ and $\{Y_j\}$. Indeed, more flexible relation among $\{Y_j\}$ could be accommodated by further specifying a joint distribution of $\{Y_j\}$ given $N$. Since the focus of this work is the association between $N$ and each $Y_j$ rather than the association within $\{Y_j\}$, we leave this potential generalization of the current model for future investigation. Second, the proposed model is flexible such that several commonly used two-part models can be viewed in the copula framework. Specific examples include the hurdle model (\cite{Mullahy1986}), the selection model (\cite{Smith2003}), and the frequency-severity model (\cite{Frees2014}). Detailed discussions can be found in Section S.1 of the supplementary material. Third, the current representation assumes $Y$ to be a nonnegative continuous outcome. However, the framework is ready to accommodate discrete outcomes with suitable modifications for (\ref{equ:pdf}). For instance, $Y$ could be a count variable in the study of health care utilization under multiple spells of illness.

\subsection{Incomplete Data} \label{sec:incomplete}

Insurance contracts typically contain some cost sharing features such as deductible and policy limit to reduce the cost of insurers. Due to such coverage modifications, $N$ and/or $Y$ are often not completely observed. Motivated by such observations, we extend the basic copula model to accommodate incomplete data.

Presumably the contract has a per-occurrence deductible $d$ and a policy limit $l$. The deductible refers to the maximal amount of loss assumed by the policyholder, and the policy limit represents the maximal possible indemnification from the insurer. Note that both quantities vary by policyholders. Given that deductible and policyholder will affect the frequency and severity observed by the insurer, we denote $\widetilde{N}$ and $\widetilde{Y}$ as the corresponding modified variables. Hence the modified aggregate loss to the insurer is:
\begin{align*}
\widetilde{S} = \widetilde{Y}_1+\cdots + \widetilde{Y}_{\widetilde{N}}.
\end{align*}

We consider two cases of incomplete data. The first one corresponds to the per-loss scenario as defined in \cite{Klugman2012}. This scenario assumes that all accidents are reported to the insurer regardless of whether the loss amount exceeds the deductbile. In this case, the frequency component is not affected by coverage modifications, thus $\widetilde{N}=N$. However, the severity component will be adjusted by:
\begin{align*}
\widetilde{Y} = \left\{
                  \begin{array}{cc}
                    0 & Y\leq d \\
                    Y-d & d<Y\leq l \\
                    l-d & Y>l \\
                  \end{array}
                \right..
\end{align*}
Thus, the joint distribution of $(\widetilde{N},\widetilde{Y}_1,\ldots,\widetilde{Y}_{\widetilde{N}})$ can be shown as:
\begin{align}
f_{\widetilde{N},\widetilde{\bm{Y}}}(n,y_1,\ldots,y_n) &= \left[f_{\widetilde{N}}(0)\right]^{I(n=0)}\left[f_{\widetilde{N}}(n)\times f_{\widetilde{\bm{Y}}|\widetilde{N}}(y_1,\ldots,y_n|n)\right]^{I(n>0)} \nonumber\\
 &= \left[f_{\widetilde{N}}(0)\right]^{I(n=0)}\left[f_{\widetilde{N}}(n)\times \prod_{j=1}^{n}f_{\widetilde{Y}|\widetilde{N}}(y_j|n)\right]^{I(n>0)} ,  \label{equ:lcoprisk}
\end{align}
where $f_{\widetilde{N}}(n) = f_N(n)$, and
\begin{align*}
f_{\widetilde{Y}|\widetilde{N}}(y|n) &= \left\{
                                         \begin{array}{cc}
                                           {\rm Pr}(\widetilde{Y}=0|\widetilde{N}=n) & y=0 \\
                                           \dfrac{\partial}{\partial y}  {\rm Pr}(\widetilde{Y}\leq y|\widetilde{N}=n) & 0<y< l-d  \\
                                           {\rm Pr}(\widetilde{Y}=l-d|\widetilde{N}=n) & y=l-d \\
                                         \end{array}
                                       \right. \\
& = \left\{
                                         \begin{array}{cc}
                                           {\rm Pr}(Y\leq d|N=n) & y=0 \\
                                           \dfrac{\partial}{\partial y}  {\rm Pr}(Y\leq y + d|N=n) & 0<y< l-d  \\
                                           {\rm Pr}(Y\ge l|N=n) & y=l-d \\
                                         \end{array}
                                       \right. \\
& = \left\{
                                         \begin{array}{cc}
                                           F_{Y|N}(d|n) & y=0 \\
                                           f_{Y|N}(y+d|n) & 0<y< l-d  \\
                                           1-F_{Y|N}(l|n) & y=l-d \\
                                         \end{array}
                                       \right..
\end{align*}
As pointed out by one reviewer, the copula between $N$ and $\widetilde{Y}$ stays unchanged since censoring is a monotone increasing function of $Y$.

The second one corresponds to the per-payment scenario as defined in \cite{Klugman2012}. Differing from the former scenario, the accident with a loss amount below the deductible is unobservable to the insurer. Hence both frequency and severity are modified by coverage modifications. The relation between the original and modified variables are:
\begin{align*}
\widetilde{N} = I(Y_1>d) +\cdots + I(Y_N>d),
\end{align*}
and
\begin{align*}
\widetilde{Y} = \left\{
                  \begin{array}{cc}
                    - & Y\leq d \\
                    Y-d & d<Y\leq l \\
                    l-d & Y>l \\
                  \end{array}
                \right.
\end{align*}

To derive the distribution of $(\widetilde{N},\widetilde{Y}_1,\ldots,\widetilde{Y}_{\widetilde{N}})$, we assume, without loss of generality, the first $k (\leq \widetilde{N}=n)$ claims are below maximum indemnification, and the rest $n-k$ claims receives maximum payments, i.e. $0<y_1,\cdots,y_k < l-d$ and $y_{k+1},\cdots,y_{n}=l-d$. Then, we have:
\begin{align}
&f_{\widetilde{N},\widetilde{\bm{Y}}}(n,y_1,\ldots,y_n)\nonumber\\
=&\dfrac{\partial ^{k}}{\partial y_1\cdots \partial y_k}{\rm Pr}(\widetilde{N}=n,\widetilde{Y}_1\leq y_1,\cdots,\widetilde{Y}_k\leq y_k,\widetilde{Y}_{k+1}=\cdots=\widetilde{Y}_n=l-d) \nonumber \\
=& {\rm E}\left[\dfrac{\partial ^{k}}{\partial y_1\cdots \partial y_k}{\rm Pr}(\widetilde{N}=n,\widetilde{Y}_1\leq y_1,\cdots,\widetilde{Y}_k\leq y_k,\widetilde{Y}_{k+1}=\cdots=\widetilde{Y}_n=l-d|N)\right] \nonumber\\
=& {\rm E}\left[\dbinom{N}{n}\dfrac{\partial ^{k}}{\partial y_1\cdots \partial y_k}{\rm Pr}(d<Y_j\leq y_j+d,j=1,\cdots,k,Y_{k+1},\cdots,Y_n>l,Y_{n+1},\cdots,Y_N\leq d|N)\right]\nonumber \\
=& {\rm E}\left[\dbinom{N}{n}\prod_{j=1}^{k}{\rm Pr}(Y_j=y_j+d)\prod_{j=k+1}^{n}{\rm Pr}(Y_j>l)\prod_{j=n+1}^{N}{\rm Pr}(Y_j\leq d)\right] \nonumber\\
=&{\rm E}\left[\dbinom{N}{n}\prod_{j=1}^k f_{Y|N}(y_j+d|N) [1-F_{Y|N}(l|N)]^{n-k} [F_{Y|N}(d|N)]^{N-n} \right] .
\end{align}

Though motivated by insurance applications, the above cases are representative of two common mechanisms for incomplete observations, censoring and truncation. Our method relies on the assumption that censoring or truncation is exogenous, i.e. the underlying distribution of $N$ and $Y$ are not affected by such mechanisms.

{\color{black}

\subsection{Inference}   \label{subsec:inference}

Because of the parametric nature of the proposed copula model}, parameters can be estimated using likelihood-based approach. Denote model parameters by $\bm\theta=(\theta^f,\theta^s,\theta^c)$, where $\theta^f$ is the vector of parameters in the frequency model, $\theta^s$ is the vector of parameters in the severity model, and $\theta^c$ represents association parameters in the bivariate copula. For complete data and censored data, one could employ either two-stage MLE or full MLE. However, for truncated data, only full MLE is available. In the following, we give detailed estimation procedures for the case of complete data. The procedures for the censored and truncated data are similar and thus omitted.

Using the basic model (\ref{equ:coprisk}), the log likelihood function for subject $i$ is shown as:
\begin{align*}
l_i(\bm{\theta}) &= \log f_N(n_i) + I(n_i>0)\times \sum_{j=1}^{n_i} \log f_{Y|N}(y_{ij}|n_i).
\end{align*}
Given a random sample $\{N_i,\bm{Y}_i\}_{i=1}^m=\{n_i,y_{i1},\ldots,y_{in_i}\}_{i=1}^m$, the full log likelihood for the case of complete data can be written as
\begin{align*}
L(\bm{\theta}) &= \sum_{i=1}^{m} \log f_N(n_i) + \sum_{\{i: n_i>0\}} \sum_{j=1}^{n_i} \log f_{Y|N}(y_{ij}|n_i) \\
& = \sum_{i=1}^{m} \log f_N(n_i) - \sum_{\{i: n_i>0\}} n_i \log f_N(n_i) \\
& ~~~~~~~~+ \sum_{\{i: n_i>0\}}\sum_{j=1}^{n_i} \left\{ \log f_{Y}(y_{ij}) + \log[h(F_N(n_i),F_Y(y_{ij}))-h(F_N(n_i-1),F_Y(y_{ij}))] \right\}.
\end{align*}

One estimation strategy is the full information likelihood method. The full MLE $\hat{\bm\theta}$ can be obtained as the maximizer of the full log likelihood function $L(\bm\theta)$. Under regularity conditions, e.g. Theorem 3.3 in \cite{NeweyMcFadden1994}, $\hat{\bm\theta}$ is consistent and asymptotically normal. The asymptotic covariance matrix of $\hat{\bm\theta}$ can be consistently estimated using the inverse of observed information at the full MLE $\hat{\bm\theta}$, i.e. $-\left[\frac{\partial^2}{\partial \bm{\theta} \partial \bm{\theta}'}L(\hat{\bm\theta})\right]^{-1}$.

The above likelihood function also suggests a two-stage estimation strategy. Denote the two stage MLE by $\hat{\bm\theta}^{2s} = (\hat{\theta}^f_{2s},\hat{\theta}^s_{2s},\hat{\theta}^c_{2s})$, and further denote:
\begin{align*}
L_1(\theta^f) &= \sum_{i=1}^m \log f_N(n_i) \\
L_2(\bm{\theta}) &= \sum_{i=1}^m I(n_i>0)\times \left[\sum_{j=1}^{n_i}\log f_{Y|N}(y_{ij}|n_i)\right],
\end{align*}
we have $L(\bm{\theta})=L_1(\theta^f)+L_2(\bm{\theta})$. In the first stage, one estimates the count regression model $f_N(n_i)$ to obtain $\hat{\theta}^f_{2s}$ by solving $\frac{\partial}{\partial \theta^f} L_1(\theta^f)=0$. Fixing the parameters in first part $\theta^f = \hat{\theta}^f_{2s}$, the second stage estimates the conditional model $f_{Y|N}(y_{ij}|n_i)$ to obtain $\hat{\theta}^s_{2s}$ and $\hat{\theta}^c_{2s}$ by solving $\frac{\partial^2}{\partial (\theta^s,\theta^c)} L_2(\hat{\theta}^f_{2s}, \theta^s, \theta^c)=0$. Under the regularity conditions of Theorem 6.1 in \cite{NeweyMcFadden1994}, $\hat{\bm\theta}^{2s}$ is consistent and asymptotically normal. However, the asymptotic covariance matrix of $\hat{\bm\theta}^{2s}$ can be tedious to calculate. The advantage of the two-stage MLE is its computational efficiency. Thus, to speed up the computation, we first obtain $\hat{\bm\theta}^{2s}$ and then use it as the initial point for the maximization of the full likelihood.

The proposed two-stage approach differs from the inference functions for marginals (IFM) method that is widely used in copula regression (\cite{Joe2005}). The IFM first estimates parameters in the univariate marginal models and then estimates the association parameters in the copula. In our case, the parameters in the severity component and the copula shall be estimated simultaneously. Applying IFM estimation to the proposed copula model will lead to inconsistent estimation because the marginal likelihood for $Y$ is not observed when $N=0$.


For model comparison, one could refer to information-based criteria such as AIC or BIC. To assess the goodness-of-fit of the copula model, we suggest the following steps. The adequacy of fit for the count regression can be examined using the standard Pearson's chi-squared test. The usual diagnostic analysis for neither the marginal distribution of $Y$ nor the bivariate copula is applicable in our case, for the same reason that the two pieces must be estimated jointly. Therefore, we employ a procedure based on the conditional distribution $f_{Y|N}$. Specifically, we calculate the fitted distribution $\widehat{F}_{Y|N}(y_{ij}|n_i)$ for $i=1,\ldots,m$ and $j=1,\ldots,n_i$. One expects the sequence $\left\{\widehat{F}_{Y|N}(y_{ij}|n_i)\right\}$ to be a sample of uniform $(0,1)$ provided that the copula model is correctly specified. In addition, one could visualize the adequacy of fit with a normal QQ plot by graphing the empirical quantiles from   $\left\{\Phi^{-1}\left(\widehat{F}_{Y|N}(y_{ij}|n_i)\right)\right\}$ against the theoretical quantiles from a standard normal distribution.
We demonstrate in detail the usage of the proposed diagnostic tools in Section S.3 of the supplementary material.

\section{Numerical Experiments}
\subsection{Impact of Dependence between $N$ and $Y$}
This section presents two numerical experiments to emphasize the importance of the dependency between $N$ and $Y$. Consider a compound distribution $S=Y_1+\cdots+Y_N$, where $N\sim Poisson(\lambda=1)$, $Y\sim Gamma(\alpha=2, \gamma=500)$, and joint distribution of $N$ and $Y$ is specified by a parametric copula. This setting is of particular interest because of the special case where $S$ is known as Tweedie compound Poisson distribution when $N$ and $Y$ are independent. As noted by \cite{Jorgensen1987}, under parameterizations $\lambda=\mu^{2-p}/[\phi(2-p)]$, $\alpha=(2-p)/(p-1)$, and $\gamma=\phi(p-1)/\mu^{p-1}$, this distribution can be expressed in the form of the exponential dispersion model with a power variance function $V(\mu)=\mu^p$ for $p \in (1,2)$.

\begin{figure}[htp]
  \begin{center}
   \includegraphics[width=0.45\textwidth,angle=270]{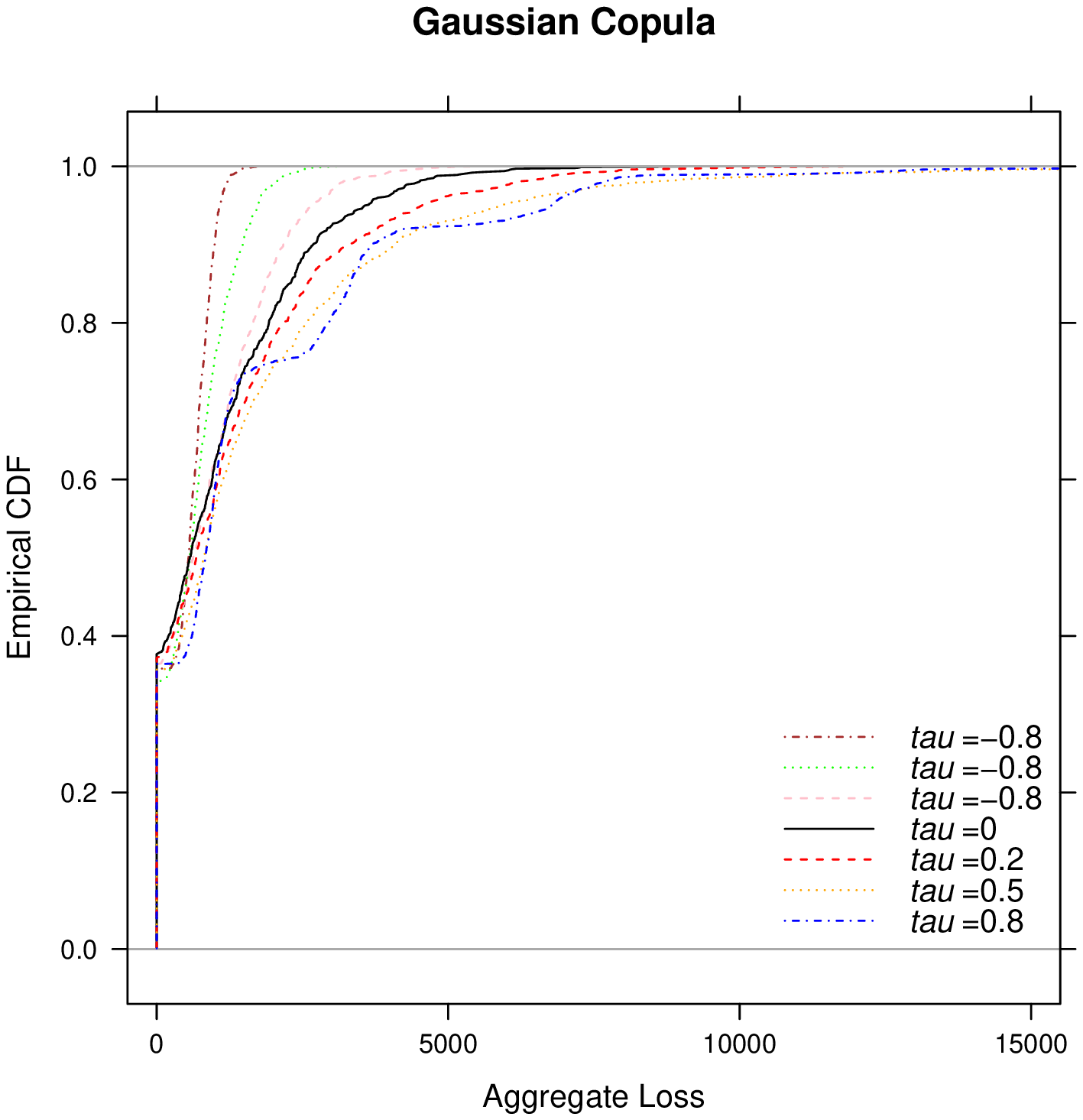}
   \includegraphics[width=0.45\textwidth,angle=270]{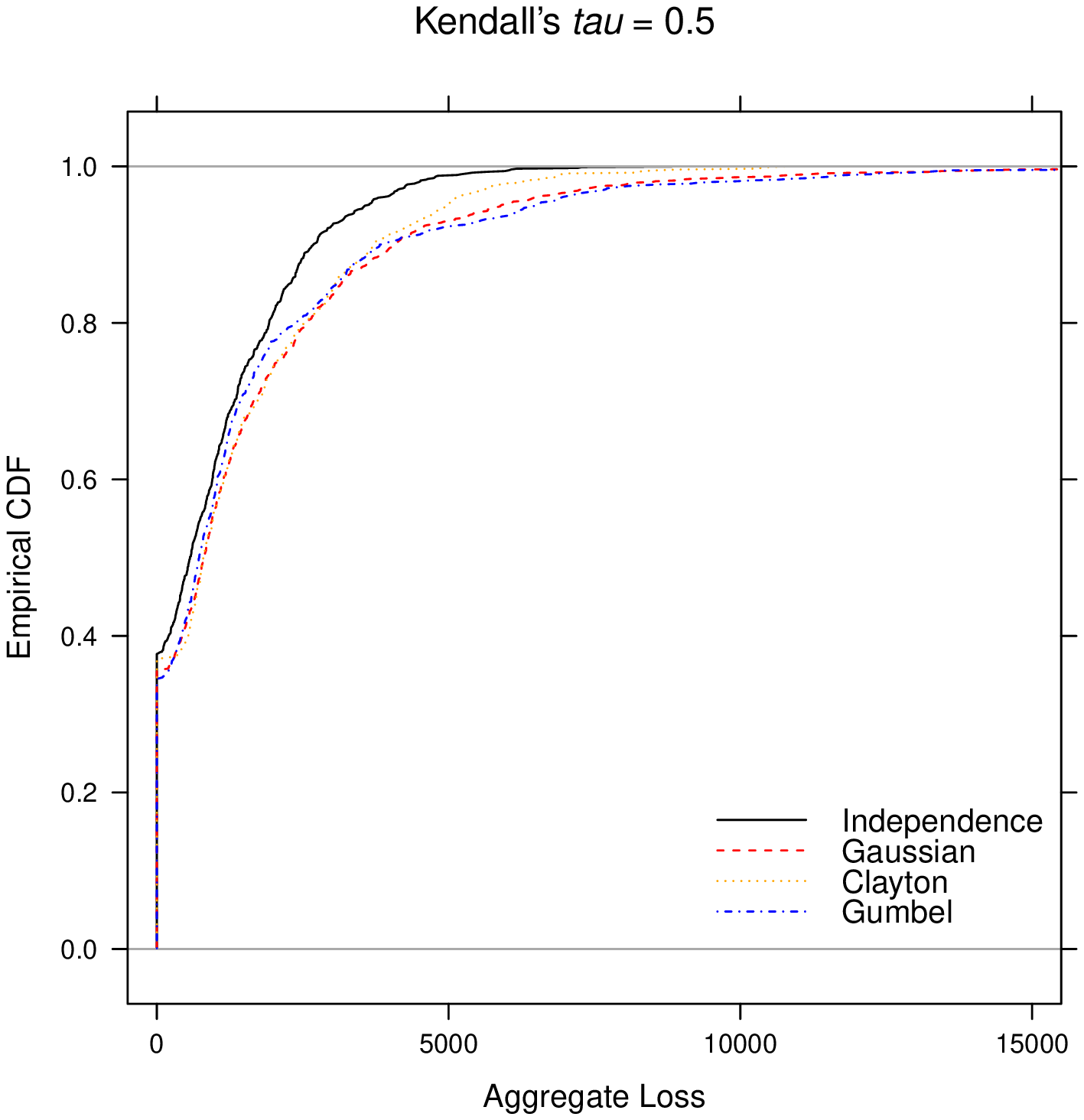}
  \end{center}
   \caption{Empirical CDF of aggregate loss. The left panel simulates data from the Gaussian copula with different Kendall's $tau$, and the right panel simulates data from different copulas with the same Kendall's $tau$.}
   \label{fig:ecdf}
\end{figure}

The first experiment demonstrates the effect of frequency-severity dependency on the distribution of aggregate loss. The distribution of $S$ is calculated using Monte Carlo simulation and is displayed in Figure \ref{fig:ecdf}. The first panel uses the Gaussian copula with different levels of dependence measured by Kendall's $tau$. When $tau=0$, the copula model reduces to the independence case which is equivalent to a Tweedie distribution ($\mu=1000,p=4/3,\phi=150$). The positive (negative) dependence leads to a longer (shorter) tail in the aggregate loss distribution. The second panel compares three copulas (Gaussian, Clayton, and Gumbel) with the same Kendall's $tau$. One observes the effect of tail dependence (upper for Gumbel and lower for Clayton), although not substantial.

\begin{figure}[htp]
  \begin{center}
   \includegraphics[width=0.32\textwidth,angle=270]{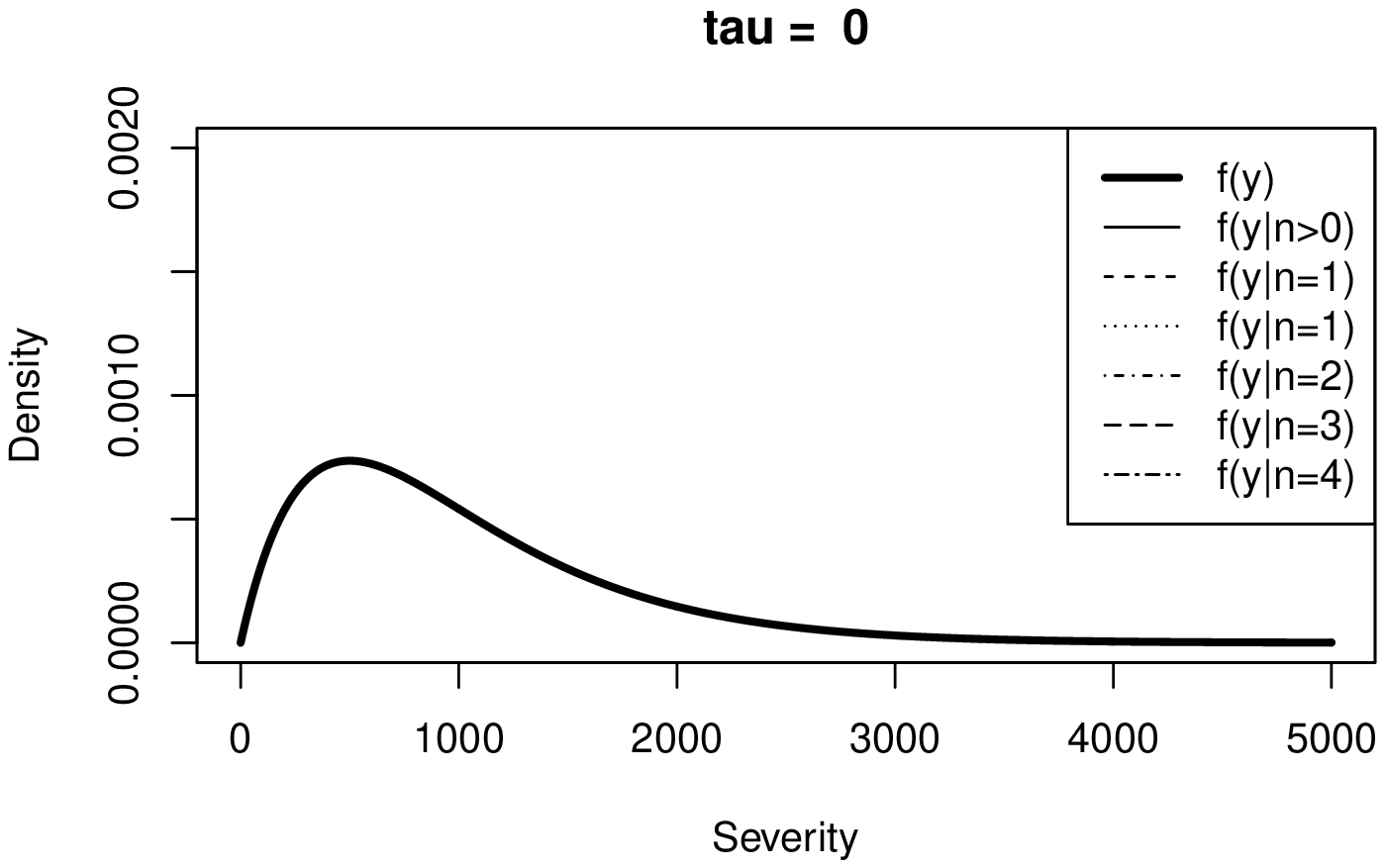}
   \includegraphics[width=0.32\textwidth,angle=270]{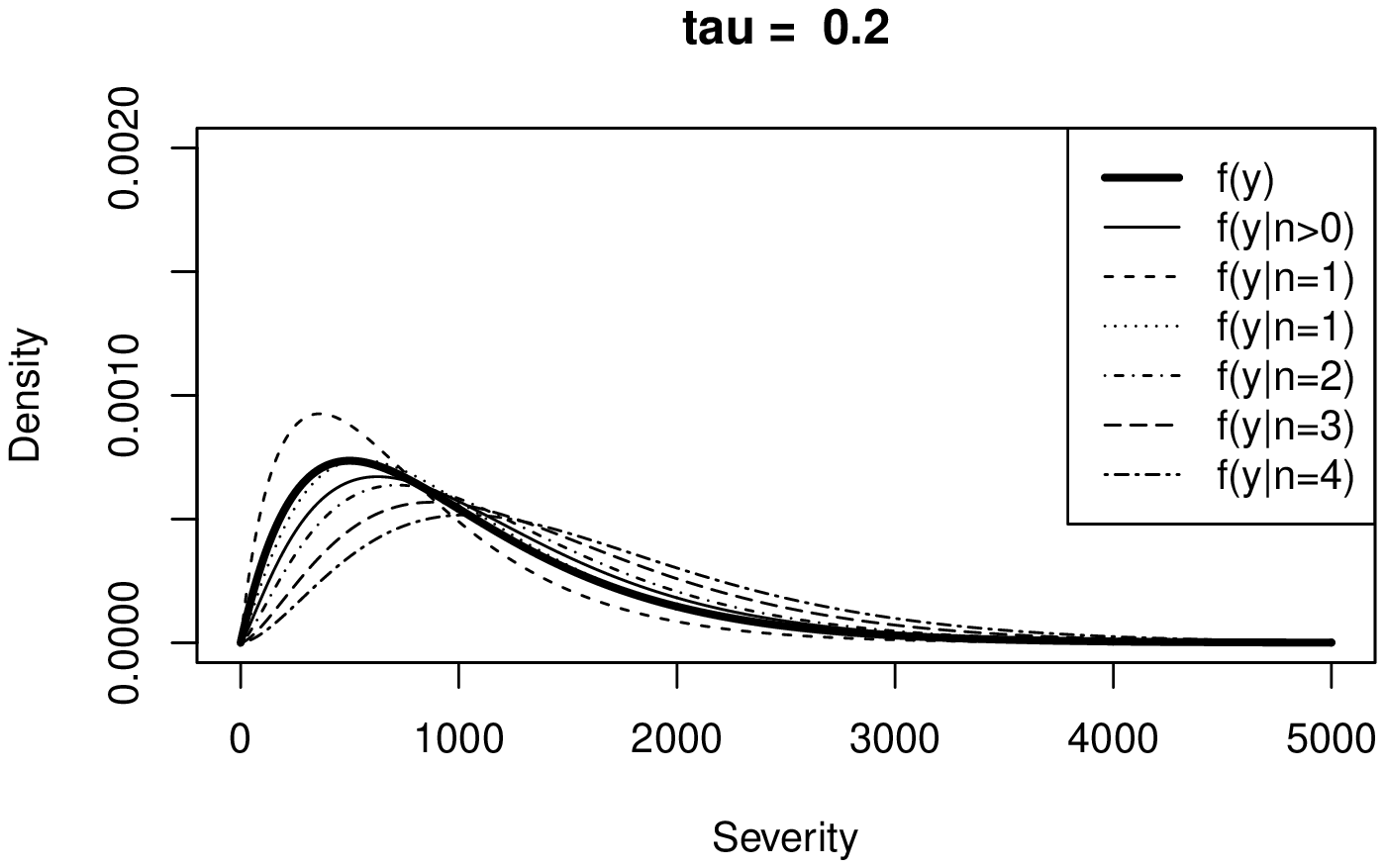}
   \includegraphics[width=0.32\textwidth,angle=270]{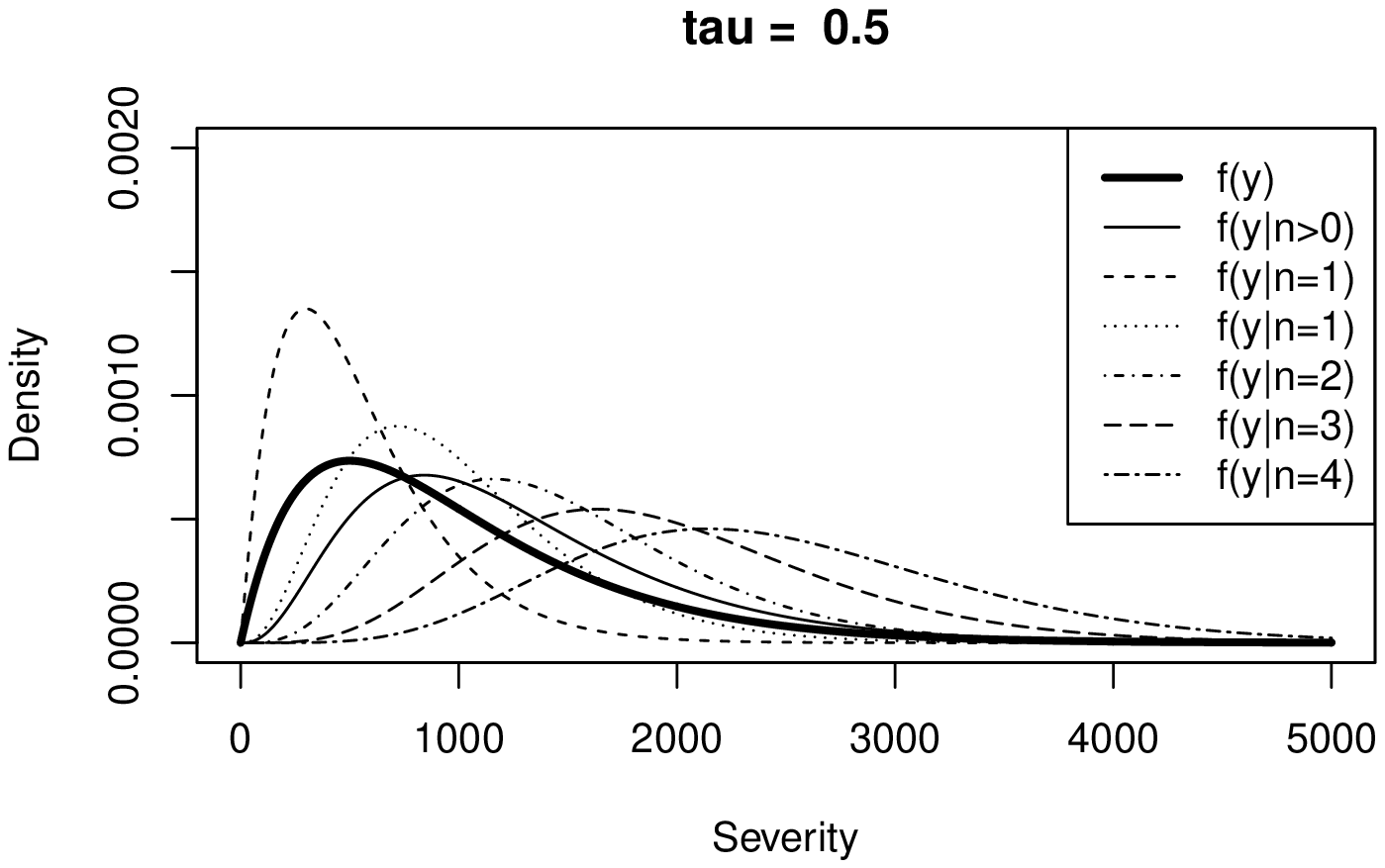}
   \includegraphics[width=0.32\textwidth,angle=270]{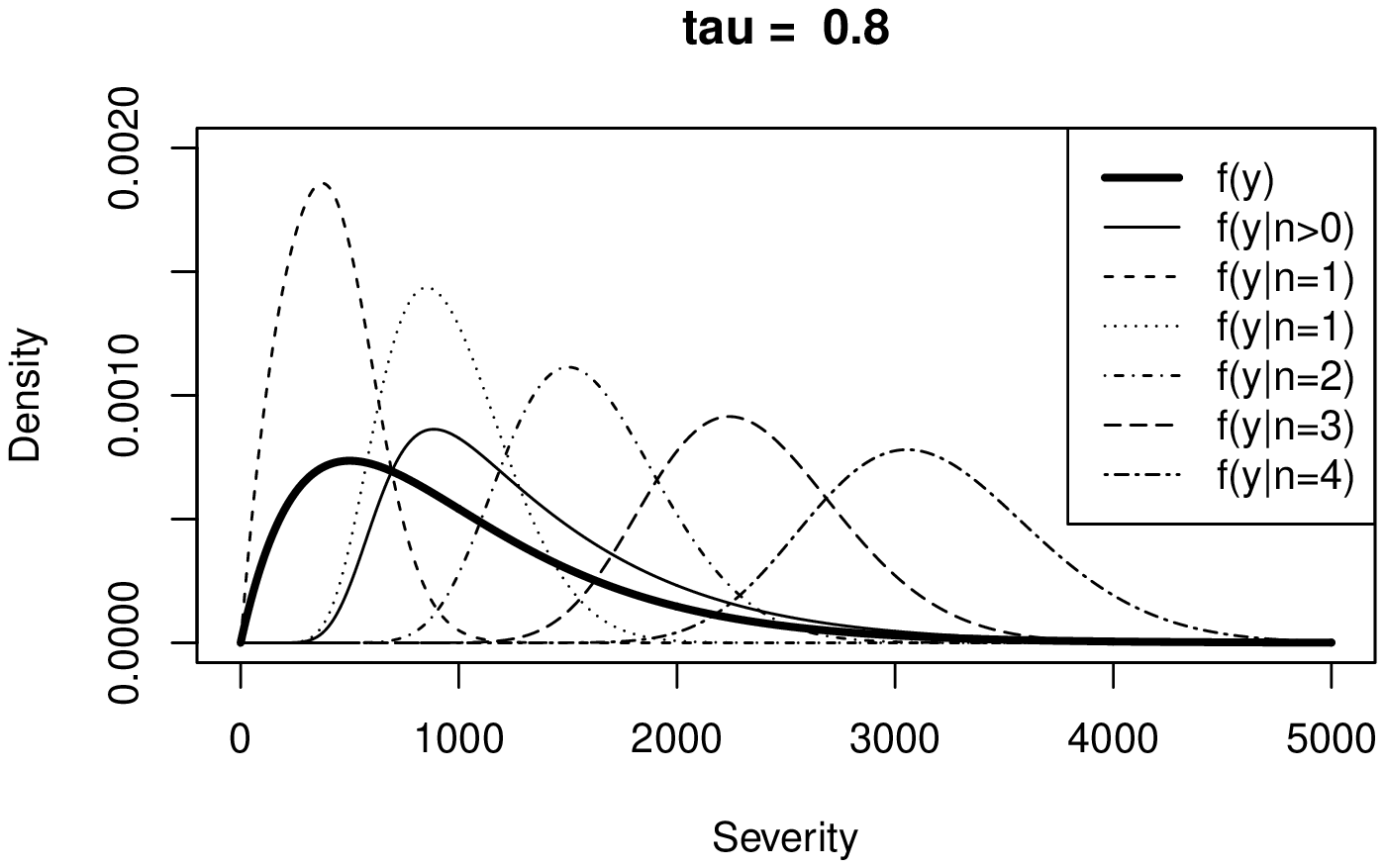}
  \end{center}
   \caption{The conditional distribution of loss amount given number of claims. The four panels correspond to different levels of dependence between claim frequency and severity.}
   \label{fig:densplot}
\end{figure}

The second experiment examines the effect of frequency-severity dependence on the conditional severity distribution. Figure \ref{fig:densplot} reports the distribution of $Y$ given $N$ at different levels of dependence. In each panel, we show densities $f_Y(y)$, $f_{Y|N>0}(y|N>0)$, and $f_{Y|N}(y|n)$. The former two cases correspond to the common practice where the claim amount is not affected by the number of claims given occurrence. The result is indicative of severe misspecification bias when the dependence between frequency and severity is ignored.

\subsection{Estimation based on the Joint Distribution of $N$ and $\bm{Y}$}
This simulation study examines the finite-sample performance of the estimations based on the joint distribution of $N$ and $\bm{Y}$, and further demonstrates the inference bias incurred by ignoring the frequency-severity dependence. We consider the Gaussian copula compound model in a regression context. The primary distribution is Poisson and the secondary distribution is gamma with:
\begin{align*}
{\rm Poisson:}&~~\log({\rm E}(N_i))=\log(\lambda_i)  = \beta_0^f + \beta_1^f X_{1i} + \beta_2^f X_{2i}\\
{\rm Gamma:}&~~\log({\rm E}(Y_{ij}))=\log(\alpha\gamma_i)  = \beta_0^s + \beta_1^s X_{1i} + \beta_2^s X_{2i},
\end{align*}
where $X_{1i}$ and $X_{2i}$ are i.i.d. and $X_1\sim Uniform(0,1)$ and $X_2\sim Bernoulli(0.5)$. In the Gaussian copula, we consider different degrees of dependence. The copula model is estimated using both the two-stage method and the joint MLE, 
and the results are summarized in Table \ref{tab:est}. We report the relative bias and the root mean squared error. The calculations are based on a sample size of 500 with 250 replications. There is no substantial difference in the estimates from the two approaches. For comparison, we also report in the table the results of the standard two-part model where $N$ and $Y$ are assumed to be independent. As anticipated, the estimates for the frequency model is consistent with the copula approach. However, the estimation assuming conditional independence introduces a long-term bias in the severity model, and this bias positively correlates with the association between $N$ and $Y$.

{\color{black}
Additional simulation studies are provided in Section S.2 of the supplementary material to illustrate the estimation for incomplete data. We emphasize that, in contrast to the cases of complete data and censored data, independence estimation will introduce persistent bias in both frequency and severity components of the model when data are truncated.}

\begin{landscape}
	\begin{table}[htbp]
		\centering
		\caption{Estimation results for complete data using the two-stage approach and the joint MLE}
		\begin{tabular}{lrrrrrrrrrrr}
			\hline\hline
			Low Dependence  &     \multicolumn{3}{c}{Independence} & &     \multicolumn{3}{c}{Two Stage} &         & \multicolumn{3}{c}{Joint MLE} \\
			\cline{2-4}\cline{6-8} \cline{10-12}
			Parameter    & Mean & Relative Bias & RMSE &  & Mean & Relative Bias & RMSE &       & Mean & Relative Bias & RMSE \\
			\hline
			$\beta_0^f$ =-1.5  & -1.515 & 0.010 & 0.107 &  & -1.515 & 0.010 & 0.107 &       & -1.518 & 0.012 & 0.114 \\
			$\beta_1^f$ = 2.5  & 2.524 & 0.009 & 0.125 & & 2.524 & 0.009 & 0.125 &       & 2.516 & 0.006 & 0.132 \\
			$\beta_2^f$ = 1    & 0.995 & -0.005 & 0.073 & & 0.995 & -0.005 & 0.073 &       & 1.002 & 0.002 & 0.075 \\
			$\beta_0^s$ = 5    & 5.092 & 0.018 & 0.124 &  & 4.988 & -0.002 & 0.093 &       & 4.991 & -0.002 & 0.091 \\
			$\beta_1^s$ = -2.5 & -2.552 & 0.021 & 0.110 &  & -2.493 & -0.003 & 0.101 &       & -2.495 & -0.002 & 0.105 \\
			$\beta_2^s$= 5     & 4.977 & -0.005 & 0.056 & & 5.004 & 0.001 & 0.054 &       & 5.001 & 0.000 & 0.051 \\
			$\alpha$ = 2     & 2.061 & 0.030 & 0.109 & & 1.998 & -0.001 & 0.097 &       & 2.005 & 0.003 & 0.093 \\
			$\rho$ = 0.1   & & &  && 0.104 & 0.039 & 0.042 &       & 0.102 & 0.023 & 0.039 \\
			\hline
			Medium Dependence &    \multicolumn{3}{c}{Independence} &  &\multicolumn{3}{c}{Two Stage} &          & \multicolumn{3}{c}{Joint MLE} \\
			\cline{2-4}\cline{6-8} \cline{10-12}
			Parameter   & Mean & Relative Bias & RMSE & & Mean & Relative Bias & RMSE &       & Mean & Relative Bias & RMSE \\
			\hline
			$\beta_0^f$ = -1.5  & -1.487 & -0.009 & 0.106 & & -1.487 & -0.009 & 0.106 &       & -1.500 & 0.000 & 0.098 \\
			$\beta_1^f$ = 2.5   & 2.478 & -0.009 & 0.118 & & 2.478 & -0.009 & 0.118 &       & 2.501 & 0.000 & 0.116 \\
			$\beta_2^f$ = 1     & 1.005 & 0.005 & 0.078 & & 1.005 & 0.005 & 0.078 &       & 0.998 & -0.002 & 0.069 \\
			$\beta_0^s$ = 5     & 5.419 & 0.084 & 0.429 & & 5.002 & 0.000 & 0.082 &       & 5.002 & 0.000 & 0.079 \\
			$\beta_1^s$ = -2.5  & -2.733 & 0.093 & 0.262 & & -2.503 & 0.001 & 0.094 &       & -2.506 & 0.002 & 0.103 \\
			$\beta_2^s$ = 5     & 4.913 & -0.017 & 0.104 & & 5.001 & 0.000 & 0.057 &       & 5.005 & 0.001 & 0.053 \\
			$\alpha$ = 2     & 2.420 & 0.210 & 0.432 &  & 2.005 & 0.003 & 0.104 &       & 2.009 & 0.004 & 0.106 \\
			$\rho$ = 0.5   & & &  && 0.501 & 0.003 & 0.028 &       & 0.500 & -0.001 & 0.026 \\
			\hline
			High Dependence&       \multicolumn{3}{c}{Independence}  &    &     \multicolumn{3}{c}{Two Stage} &        & \multicolumn{3}{c}{Joint MLE} \\
			\cline{2-4}\cline{6-8} \cline{10-12}
			Parameter   & Mean & Relative Bias & RMSE &  & Mean & Relative Bias & RMSE &       & Mean & Relative Bias & RMSE \\
			\hline
			$\beta_0^f$ =-1.5  & -1.509 & 0.006 & 0.110 & & -1.509 & 0.006 & 0.110 &       & -1.503 & 0.002 & 0.078 \\
			$\beta_1^f$ =2.5   & 2.507 & 0.003 & 0.134 & & 2.507 & 0.003 & 0.134 &       & 2.497 & -0.001 & 0.091 \\
			$\beta_2^f$ = 1    & 1.003 & 0.003 & 0.080 & & 1.003 & 0.003 & 0.080 &       & 1.002 & 0.002 & 0.058 \\
			$\beta_0^s$ = 5     & 5.690 & 0.138 & 0.698 & & 4.999 & 0.000 & 0.090 &       & 5.000 & 0.000 & 0.058 \\
			$\beta_1^s$ = -2.5  & -2.870 & 0.148 & 0.395 & & -2.500 & 0.000 & 0.129 &       & -2.507 & 0.003 & 0.082 \\
			$\beta_2^s$ = 5     & 4.855 & -0.029 & 0.166 & & 5.001 & 0.000 & 0.069 &       & 5.001 & 0.000 & 0.050 \\
			$\alpha$ = 2     & 3.083 & 0.541 & 1.109 & & 2.004 & 0.002 & 0.081 &       & 2.000 & 0.000 & 0.077 \\
			$\rho$ = 0.9   & & &  && 0.900 & 0.000 & 0.006 &       & 0.901 & 0.001 & 0.006 \\
			\hline\hline
		\end{tabular}%
		\label{tab:est}%
	\end{table}%
\end{landscape}

\section{Modeling Aggregate Insurance Claims}
In nonlife insurance (including property, casualty, and health), the compound distribution (\ref{equ:crm}) is a common approach to modeling aggregate losses in an insurance system. Examples of an insurance system include a single policyholder, a line of business, or a portfolio of contracts. The compound distribution is known as collective risk model in the actuarial literature, and the frequency and severity components are the two building blocks of the model (\cite{Klugman2012}).

In this application, we examine the Wisconsin local government property fund which provides property insurance for local government entities in the state of Wisconsin, such as court houses, school districts, fire stations, etc. We consider the building and contents coverage where the building element covers for the physical structure of a property including its permanent fixtures and fittings, and the contents element covers possessions and valuables within the property that are detached and removable. Similar to most nonlife insurance product, the contract provided by the property fund has a one-year term.

The insurance system in this context corresponds to a policyholder, that is a local government entity. The outcome of interest is the aggregate loss for an entity during the policy year, which is determined by both the number and the size of claims. {\color{black} As discussed in Section \ref{sec:intro},} the collective risk model implies a frequency-severity approach for modeling the aggregate loss for each policyholder, and the current practice relies on the independence assumption between the two building blocks $N$ and $Y$ in the collective risk model.

The purpose of the analysis below is two-folded: first, we provide empirical evidence of significant negative association between the frequency and severity of insurance claims; second, we show that ignoring the frequency-severity dependence could lead to biased decision-making in insurance operations. In the following sections, we use term ``independence model'' to refer to the standard frequency-severity model that assumes independence between the frequency and severity components, {\color{black} and ``copula model'' to refer to the proposed copula approach in Section \ref{sec:copulamodel}.1 that allows for flexible dependence between the frequency and severity components.}

Granular insurance claim data are collected for a portfolio of local government entities for years 2009-2011. For each policyholder, one observes the number of claims and the ground-up loss of each claim during each year. We use data of 2009 and 2010 to develop the model, and data of 2011 for model validation. There are 2,080 and 1,017 policy-year observations in the training data and validation data, respectively. 

\subsection{Exploring Frequency-Severity Association} \label{subsec:relationexplore}

\begin{figure}[htp]
  \begin{center}
   \includegraphics[width=0.6\textwidth,angle=270]{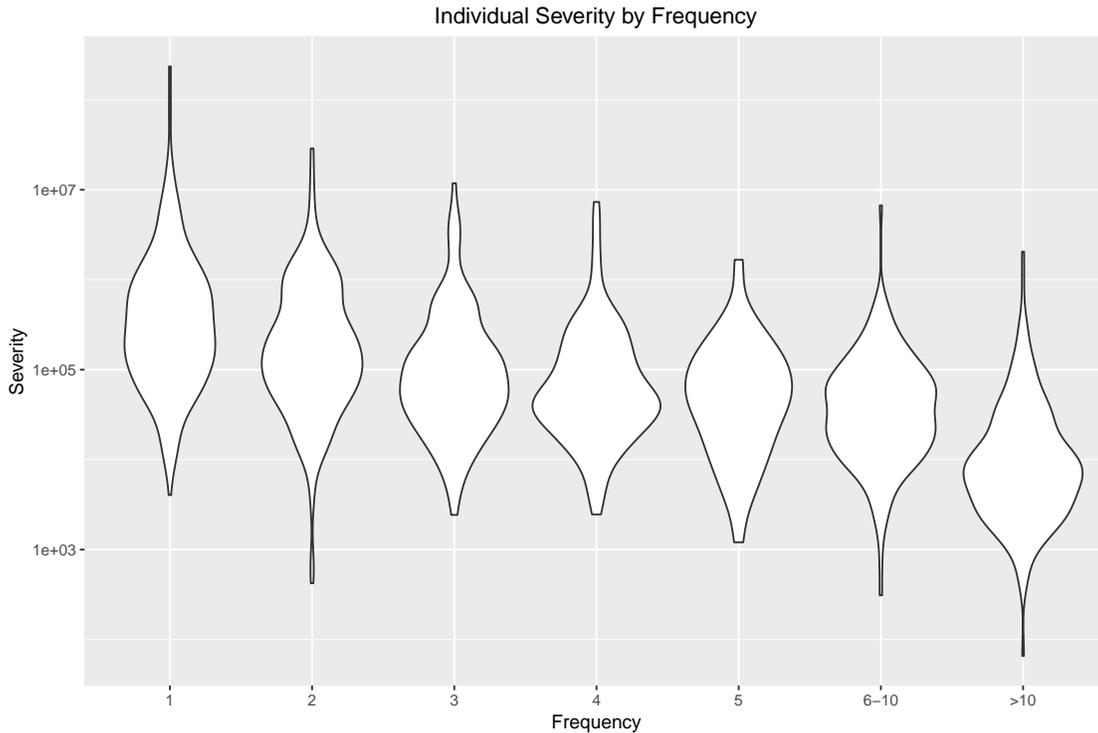}
  \end{center}
   \caption{Violin plot of claim amount per \$1,000 coverage by the number of claims.}
   \label{fig:sevbyfreq}
\end{figure}

To explore the relationship between claim frequency and severity, we display in Figure \ref{fig:sevbyfreq} the violin plot of claim size by the number of claims for the portfolio of government entities. To account for exposure, the claim size is normalized by the amount of coverage. First, one observes that given occurrence, the distribution of claim severity correlates with claim frequency. Second, the violin plot suggests a negative relation between claim severity and frequency, i.e. the amount of claims tends to be smaller for policyholders who have more claims.

To further motivate the usage of the proposed copula model, we perform some preliminary analyses to examine the role of frequency-severity dependence in model fitting. Our starting point is the Tweedie model given it is the industry standard in property-casualty insurance for modeling semi-continuous loss cost. Recall that the Tweedie distribution is a Poisson sum of gamma variables where the Poisson and gamma variables are assumed to be independent. To examine the role of dependence, we further allow the Possion and gamma variables in the Tweedie distribution to be correlated. Specifically, we fit a copula model for the aggregate loss where the frequency is a Poisson variable, and the severity is a gamma variable, and their joint distribution is specified by a bivariate Gaussian copula. The association parameter in the Gaussian copula is estimated to be $-0.278$ with a standard error of $0.022$. This result is consistent with the pattern suggested by Figure \ref{fig:sevbyfreq}.

\begin{figure}[htp]
	\begin{center}
		\includegraphics[width=0.4\textwidth,angle=270]{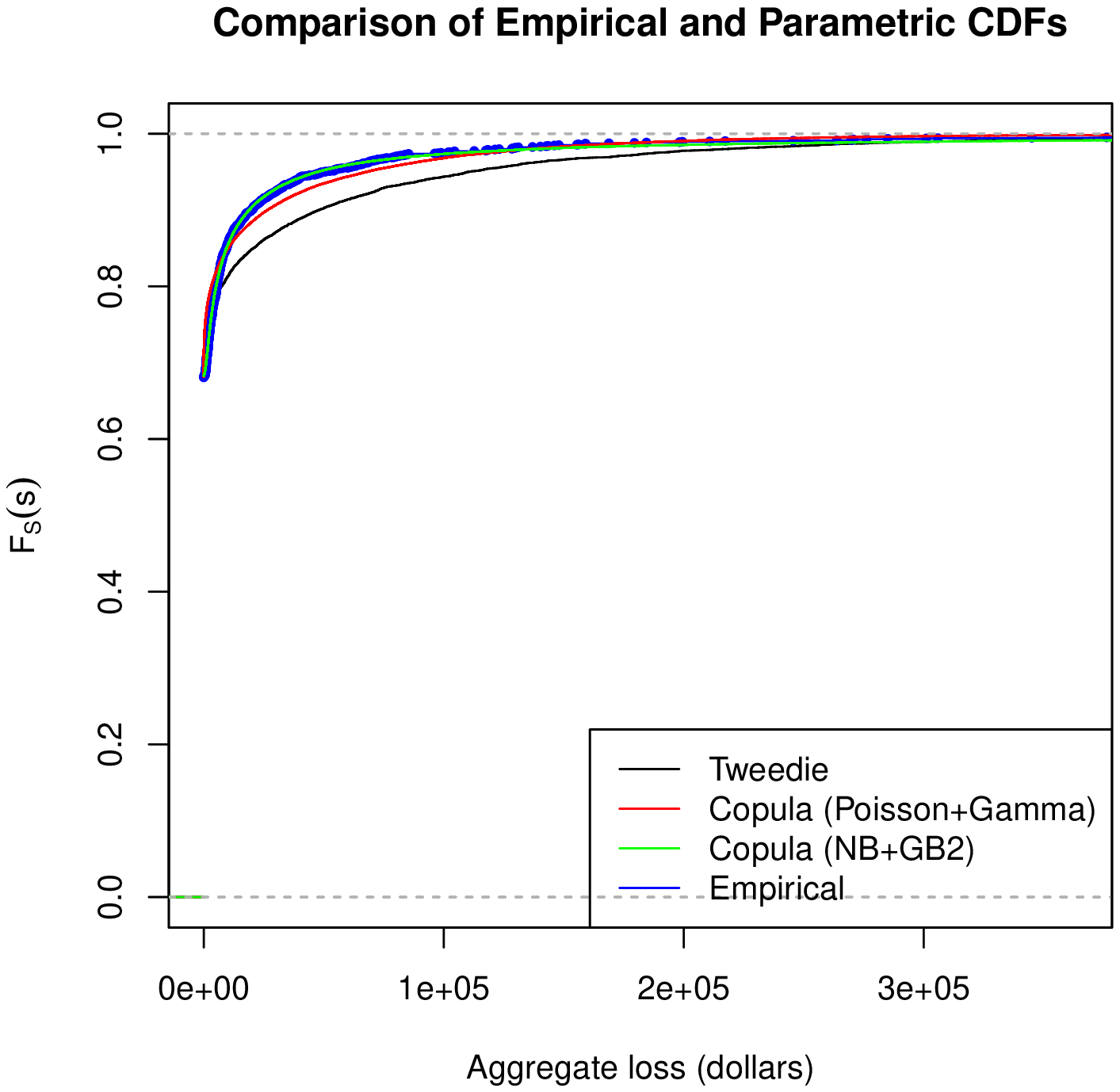}
		\includegraphics[width=0.4\textwidth,angle=270]{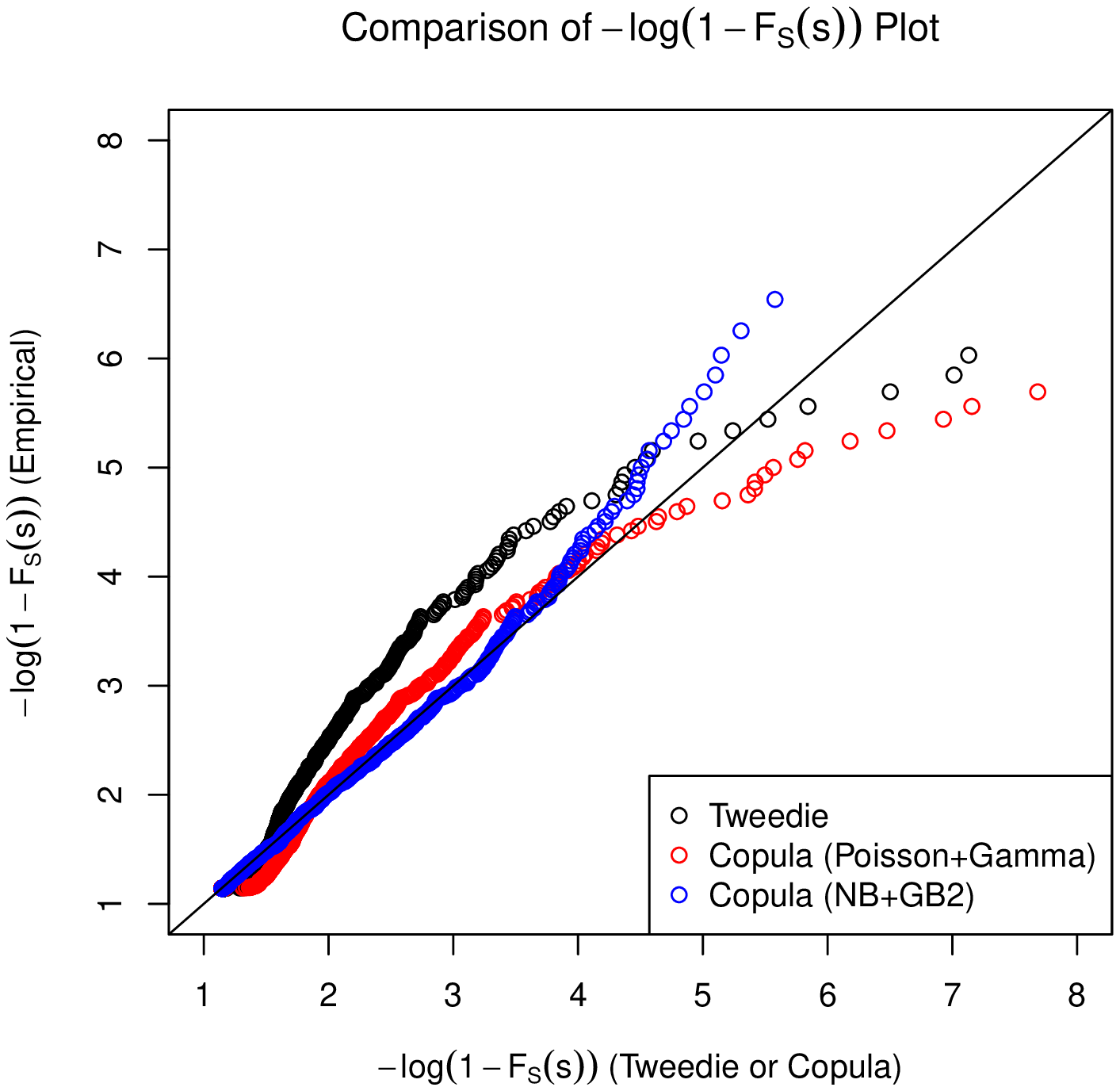}
	\end{center}
	\caption{Comparison between empirical and parametric Cumulative Distribution Functions (CDF, denoted by $F_S(s)$) of aggregate loss.}
	\label{fig:ecdf1}
\end{figure}

To compare the Tweedie and copula models, we present in Figure \ref{fig:ecdf1} two goodness-of-fit plots. Denote $F_S(s)$ as the cumulative distribution function~(CDF) of aggregate loss. The left figure shows the fitted CDF of the aggregate loss from the two parametric models along with the empirical estimate. Since the plot of CDF emphasizes the center of the distribution, it is not ideal to visualize the effects of extremal large values. To further investigate the tail fit, the right figure plots $-\log(1-F_S(s))$ between the empirical distribution and the two parametric (Tweedie and copula) models. {\color{black} On one hand, both plots indicate that the copula model exhibits superior fit to the Tweedie model, emphasizing the importance of frequency-severity dependence. On the other hand, there is still room for improvement of goodness-of-fit in both the center and the tail of the distribution. This suggests considering more flexible distributions for marginal behavior. To illustrate, we fit another copula model using zero-one inflated negative binomial distribution for claim frequency, the generalized beta of the second kind (GB2) distribution for claim severity, and a Gaussian copula between the two components. The estimated association parameter is $-0.207$ with a standard error of $0.032$. The corresponding goodness-of-fit plots are also shown in Figure \ref{fig:ecdf1}. As anticipated, refined marginal models improve the fit, especially in the heavy right tail.} Overall, the preliminary analyses suggests that there is significant negative dependence between claim frequency and severity, and accounting for such association enhances the goodness-of-fit for the aggregate loss distribution.

\subsection{Empirical Analysis}

\textcolor{black}{The observation in Section \ref{subsec:relationexplore} motivates us to jointly examine the frequency and severity components in the collective risk model.} Differing from the earlier preliminary analysis, first, we explore using more flexible marginal distributions for modeling the number and the size of insurance claims; second, we incorporate covariates to account for observed heterogeneity, and thus the relation between frequency and severity is interpreted as residual dependence; {\color{black} third, we consider various copula that offer different types of dependence in modeling the frequency-severity relationship.}

\begin{figure}[htp]
  \begin{center}
   \includegraphics[width=0.5\textwidth,angle=270]{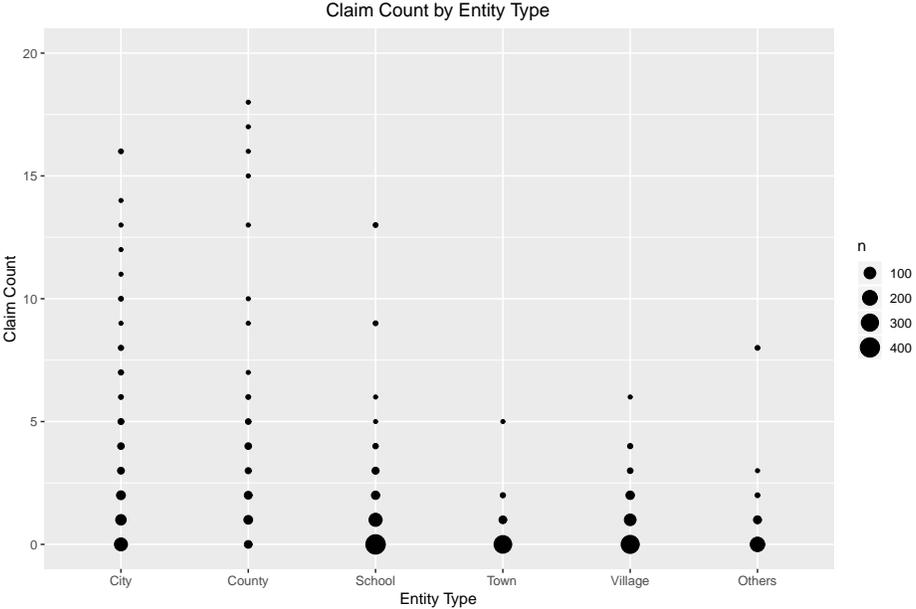}
  \end{center}
   \caption{Distribution of claim count by entity type.}
   \label{fig:countentity}
\end{figure}

To facilitate model specification, we examine the distributions of both claim frequency and severity, as well as their relationship with available explanatory variables. The insurance database contains policyholder-specific and claim-specific information that one could use to account for the variation in claim frequency and severity. {\color{black} Details of such covariate information are provided in Section S.3 of the supplementary material.} For claim frequency, we consider the policy-level characteristics, including entity type (whether a policyholder is a city, county, township, village, or others), alarm credit (whether a policyholder receives a credit for alarm system), the level of deductible, and the amount of coverage. For illustration, we exhibit in Table \ref{tab:freqtable} the empirical distribution of the number of claims per policyholder in the training data. As usually observed in insurance claims data, the majority of policyholders (about 70\%) has zero claims over the year. However, this percentage is much smaller than private lines of business such as personal automobile insurance. Another striking feature of claim counts is there is an excess of ones in addition to the zero inflation. We further break down the frequency distribution by entity type, as shown in Table \ref{tab:freqtable} and visualized in Figure \ref{fig:countentity}. The substantial variation suggests that entity type is an important predictor for the claim count.


\begin{table}[htbp]
	\centering
	\caption{Distribution of claim frequency: overall and by entity type (in percentage)}
\begin{tabular}{crrrrrrr}
\hline\hline
     &    &       \multicolumn{6}{c}{Entity Type} \\
     \cline{3-8}
        Frequency   &    Overall &       City &     County &     School &       Town &    Village &     Others \\
\hline
         0 &      68.08 &      45.67 &      19.67 &      67.11 &      91.95 &      70.33 &      85.45 \\
         1 &      19.38 &      24.00 &      31.15 &      23.36 &       6.90 &      20.75 &      12.27 \\
         2 &       6.54 &      13.33 &      20.49 &       5.26 &       0.86 &       7.05 &       0.91 \\
         3 &       2.12 &       4.67 &       6.56 &       2.63 &       0.00 &       1.04 &       0.45 \\
         4 &       1.49 &       4.00 &       9.84 &       0.66 &       0.00 &       0.62 &       0.00 \\
         5 &       0.67 &       2.33 &       4.10 &       0.16 &       0.29 &       0.00 &       0.00 \\
       $\ge 6$ &       1.73 &       6.00 &       8.20 &       0.82 &       0.00 &       0.21 &       0.91 \\
 \hline
      Obs  &       2080 &        300 &        122 &        608 &        348 &        482 &        220 \\
\hline\hline
\end{tabular}
	\label{tab:freqtable}%
\end{table}%

\begin{figure}[htp]
  \begin{center}
   \includegraphics[width=0.45\textwidth,angle=270]{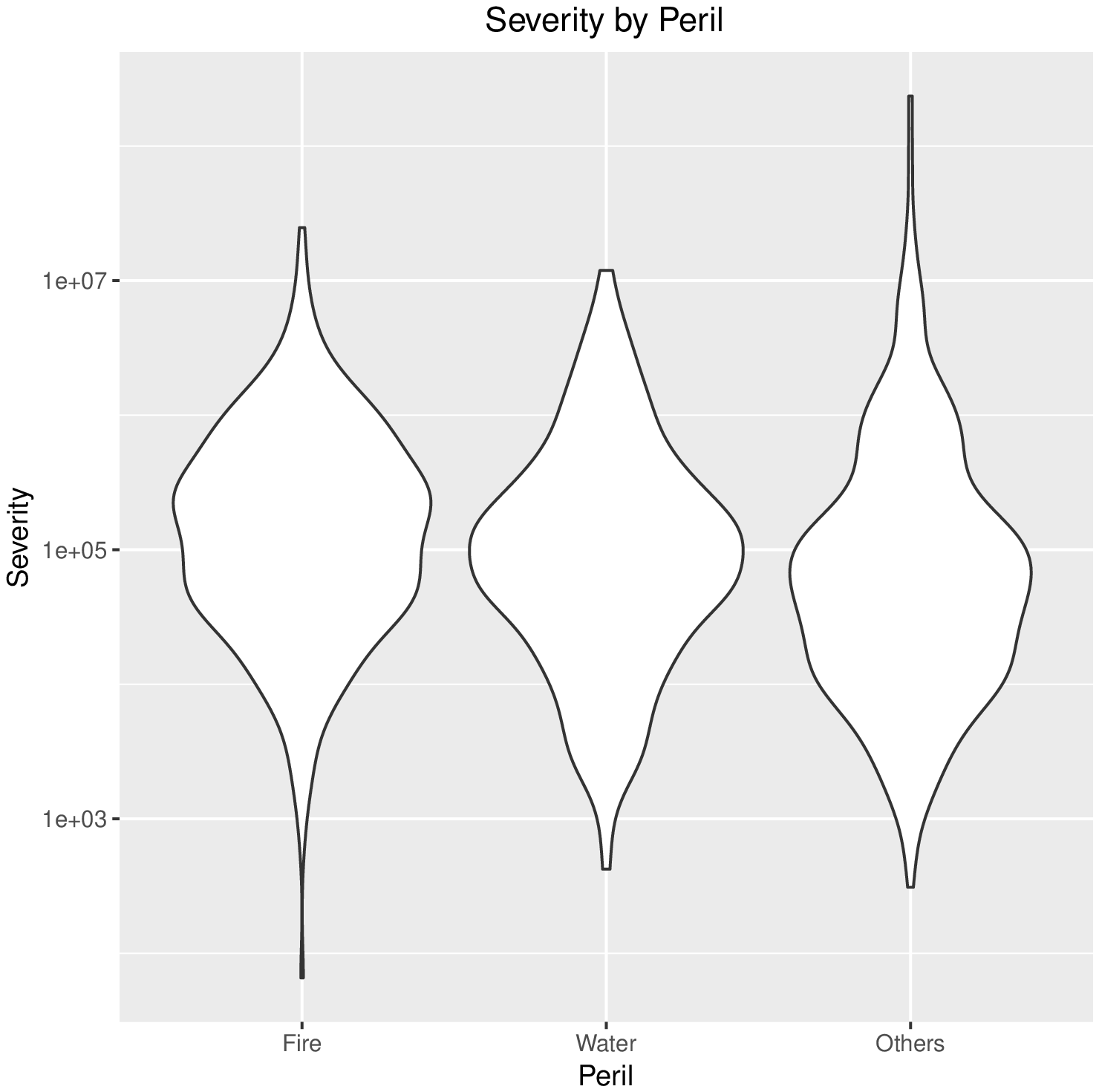}
   \includegraphics[width=0.45\textwidth,angle=270]{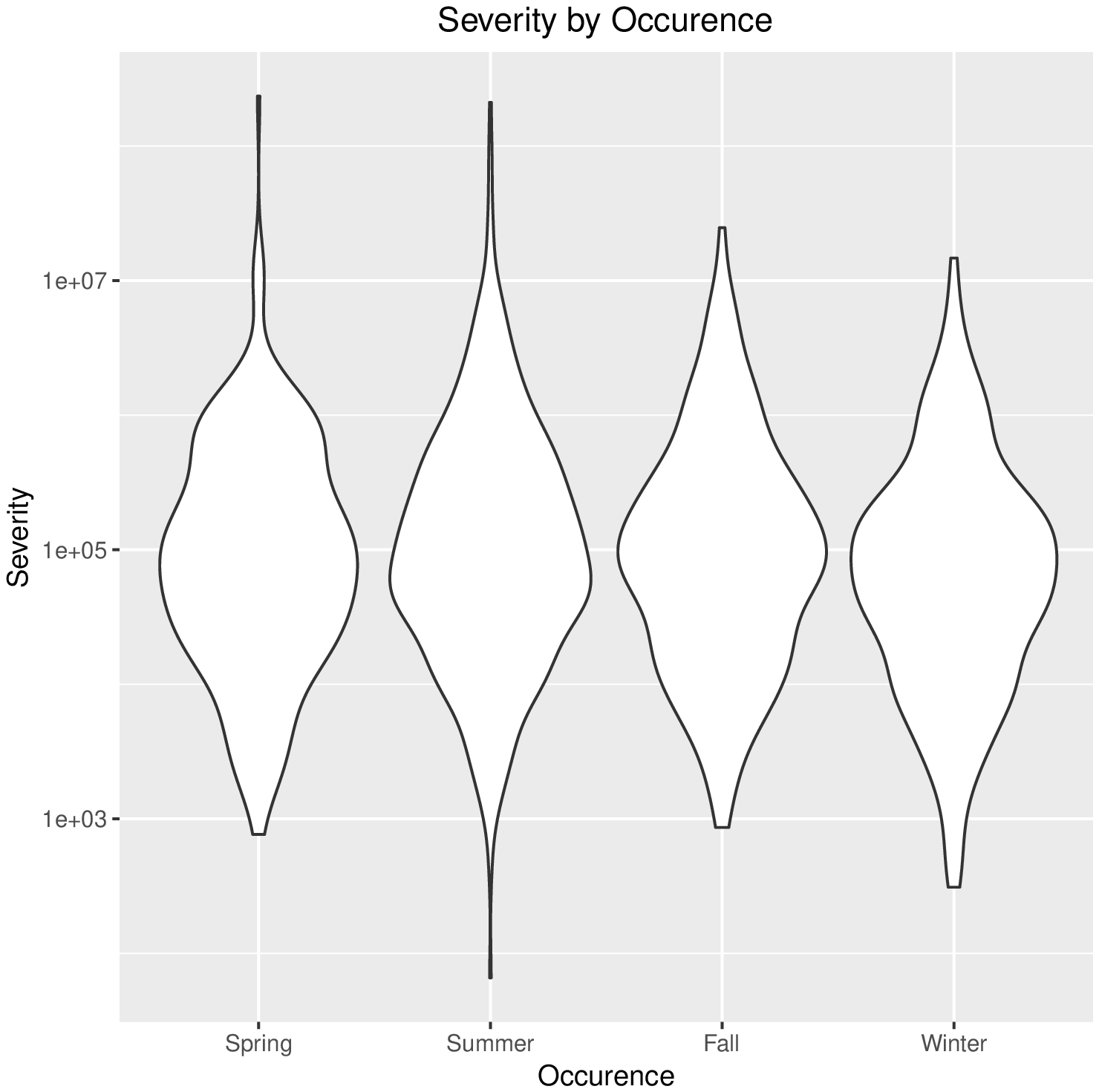}
  \end{center}
   \caption{Violin plots of claim severity. The left and right panels show severity distributions by peril and occurrence respectively.}
   \label{fig:sevperil}
\end{figure}

Table \ref{tab:severitytable} summarizes the empirical quantiles of claim amounts. There are in total 1,381 claims in the sampling period. The descriptive statistics indicates that claim amount is skewed and heavy-tailed distributed. For claim severity, besides policy-level information, we look into the effects of claim-level information such as peril type, occurrence time, and reporting delay. As an example, Table \ref{tab:severitytable} shows the empirical distribution of claim amount by peril type and by occurrence time. The claim amount due to fire and water damages tends to be larger compared to other perils, and the loss events occurred in the summer is more likely to result in higher claims. The pattern is also displayed in the violin plot of the claim severity in log scale in Figure \ref{fig:sevperil}. The plot reinforces the skewness in the severity distribution and stresses the heterogeneity across occurrence and peril type. 

\begin{table}[htbp]
  \centering
  \caption{Distribution of claim amount: overall, by peril, and by occurrence (in dollars)}
    \begin{tabular}{crrrrrrrrr}
    \hline\hline
     &    &       \multicolumn{3}{c}{Peril}&  & \multicolumn{4}{c}{Occurrence}  \\
\cline{3-5} \cline{7-10}
    Quantiles & Overall & Fire & Water & Others && Spring & Summer & Fall &Winter\\
        \hline
    10 & 946   & 1,072 & 1,009 & 790    && 991   & 950   & 945   & 912 \\
    25 & 1,645 & 2,168 & 1,641 & 1,418  && 1,600 & 1,655 & 1,746 & 1,666 \\
    50 & 3,542 & 4,989 & 4,200 & 2,945  && 3,021 & 3,859 & 3,802 & 3,619 \\
    75 & 9,062 & 13,069 & 11,305 & 5,724  && 7,219 & 11,838 & 8,852 & 7,155 \\
    90 & 29,288 & 29,849 & 35,640 & 22,203  && 27,872 & 34,181 & 26,890 & 26,758 \\
    \hline
    Obs   & 1381  & 400   & 389   & 592    && 290   & 539   & 289   & 263 \\
    \hline\hline
    \end{tabular}%
  \label{tab:severitytable}%
\end{table}%

In the final model, we consider a zero-one inflated negative binomial regression for claim frequency:
\begin{align} \label{equ:mcount}
f_{N}(n_i) = p_{i}^{0}I(n_i=0) + p_{i}^{1} I(n_i=1) + (1-p_{i}^{0}-p_{i}^{1})g_{N}(n_i),
\end{align}
where $p_{i}^{k}$ ($k=0,1$) is specified using a multinomial logistic regression:
\begin{align*}
p_{i}^{k} = \frac{\exp(\bm{x}'_{i}\bm{\beta}^f_k)}{1+ \sum_{k=0}^1\exp(\bm{x}'_{i}\bm{\beta}^f_k)},~~k=0,1,
\end{align*}
and $g_{N}(\cdot)$ is a standard negative binomial model:
\begin{align*}
g_{N}(n_i) = \frac{\Gamma(\eta+n_i)}{\Gamma(\eta)\Gamma(n_i+1)}\left[\frac{\eta}{\eta+\exp(\bm{x}_i\bm{\beta}^f)}\right]^{\eta}\left[\frac{\exp(\bm{x}_i\bm{\beta}^f)}{\eta+\exp(\bm{x}_i\bm{\beta}^f)}\right]^{n_i},
\end{align*}
with $\eta>0$ being the dispersion parameter. This specification allows to accommodate the excess of both zeros and ones in the claim count. To accommodate the skewness and heavy-tails, a parametric regression based on GB2 distribution is employed for claim severity (for instance, see \cite{Shi2014} for details on GB2 regression):
\begin{align} \label{equ:mseverity}
f_{Y}(y_{ij}) = \cfrac{[\exp(w_{ij})]^{\phi_1}}{y_{ij}|\sigma|B(\phi_1,\phi_2)[1+\exp(w_{ij})]^{\phi_2}},
\end{align}
where $\phi_1$ and $\phi_2$ are shape parameters, $\sigma$ is the scale parameter, and $w_{ij}=(\log y_{ij} - \bm{x}'_i\bm{\beta}^s)/\sigma$. A parametric bivariate copula is employed to construct the joint distribution of $N$ and $Y$. We consider commonly used bivariate copulas from the elliptical and Archimedean families, including Gaussian, $t$, Clayton, Frank, Gumbel, and Joe. For the Archimedean copulas that only allow for positive association, we consider the associated 90 and 270 degree rotated copulas.

{\color{black} The copulas models are estimated using likelihood-based estimation described in Section \ref{subsec:inference}.} The corresponding goodness-of-fit statistics are reported in Table \ref{tab:aic}. The independence model is presented as a benchmark. Model selection criteria AIC and BIC recommend the Gaussian copula model. It appears that the tail dependence is not a concern in this context. The implied Kendall's $tau$, reported in the table, reinforces the negative frequency-severity dependence obtained in the earlier analysis, indicating that the claim frequency and severity are correlated after controlling for the covariates. Because the independence model is nested by the copula model, we perform a likelihood ratio test to formally evaluate the goodness-of-fit of the copula models against the independence model. The large $\chi^2$ statistics confirm the statistical significance of the negative frequency-severity dependence.

The specification for the dependent frequency-severity model, including both the marginals and the copula, is a result of a series of model comparisons, diagnostic analysis, and robust checks. The detailed analysis is provided in Section S.3 of the supplementary material.


\begin{table}[htbp]
  \centering
  \caption{Goodness-of-fit statistics for various copula models}
    \begin{tabular}{lrrrrr}
    \hline\hline
          & {Kendall's $tau$} & {LogLik} & {AIC} & {BIC} & {Pearson's $\chi^2$} \\
    \hline
    Independence &       & -15,756 & 31,587 & 31,801 &  \\
    Gaussian & -0.19 & -15,720 & 31,518 & 31,738 & 70.77 \\
    $t$     & -0.19 & -15,719 & 31,519 & 31,744 & 72.38 \\
    Clayton90 & -0.08 & -15,723 & 31,523 & 31,743 & 66.06 \\
    Clayton270 & -0.33 & -15,739 & 31,555 & 31,775 & 34.09 \\
    Gumbel90 & -0.29 & -15,722 & 31,521 & 31,741 & 68.04 \\
    Gumbel270 & -0.09 & -15,731 & 31,540 & 31,759 & 49.53 \\
    Frank90/270 & -0.22 & -15,733 & 31,544 & 31,764 & 45.49 \\
    Joe90 & -0.34 & -15,739 & 31,557 & 31,777 & 32.38 \\
    Joe270 & -0.05 & -15,735 & 31,548 & 31,768 & 41.41 \\
    \hline\hline
    \end{tabular}%
  \label{tab:aic}%
\end{table}%

Table \ref{tab:modelest} reports the estimated parameters for the selected Gaussian copula model. The association parameter in the Gaussian copula is -0.29 and -0.30 using two-stage and full MLE respectively. {\color{black} Given that the rating variables in insurance are highly regulated, one should regard the observed frequency-severity dependence as a result of unobserved heterogeneity, and thus the sign of the dependence could be either positive and negative. Our focus is to provide a data-driven method to capture such relationship and to show the detrimental effects of ignorant supposition of independence on statistical inference and hence insurance operations.} For comparison, we also report in Table \ref{tab:modelest} the estimation results for the independence model. For the frequency component, one anticipates no essential difference in estimates of regression coefficients between the independence and copula models. We observed that the two-stage MLE is identical to the independence model, and we attribute the difference from the full MLE to the finite sample property.  In contrast, the difference in the estimates for the severity component is substantial between the independence and copula models~(both two-stage and full MLE), which is in line with the significant negative dependence between $N$ and $Y$. The analysis indicates that ignoring the frequency-severity dependence could introduce significant bias in parameter estimation.

\begin{landscape}
\begin{table}[htbp]
  \centering
  \caption{Parameter estimation for the independence model and the copula model}
    \begin{tabular}{lrrrrrrrrrrrrrrrrr}
    \hline\hline
     & \multicolumn{5}{c}{Independence} & & \multicolumn{5}{c}{Copula-Two Stage MLE}&  & \multicolumn{5}{c}{Copula-Full MLE} \\
     \cline{2-6} \cline{8-12} \cline{14-18}
     &\multicolumn{2}{c}{Frequency}  & &\multicolumn{2}{c}{Severity} & &\multicolumn{2}{c}{Frequency}  &&\multicolumn{2}{c}{Severity}  & &\multicolumn{2}{c}{Frequency}  &&\multicolumn{2}{c}{Severity}\\
     \cline{2-12}
          & Est. & S.E. &       & Est. & S.E. &       & Est. & S.E. &       & Est. & S.E.&    & Est. & S.E. &       & Est. & S.E.\\
    \hline
    Intercept & -1.184 & 0.375 &       & 7.212 & 0.289 &       & -1.184 & 0.377 &       & 7.031 & 0.317 &   & -0.728 &  0.397 &       &  6.886 & 0.324 \\
    City  & 0.299 & 0.257 &       & -0.333 & 0.198 &     & 0.299 & 0.244 &       & -0.548 & 0.216 &    & 0.485 & 0.232 &       & -0.616 & 0.216 \\
    County & 0.169 & 0.285 &       & -0.352 & 0.209 &       & 0.169 & 0.269 &       & -0.637 & 0.231&    & 0.391 & 0.260 &       & -0.717 & 0.232 \\
    School & -0.872 & 0.262 &       & 0.141 & 0.205 &        & -0.872 & 0.250 &       & 0.027 & 0.221&  &  -0.636 & 0.238 &       & -0.055 & 0.222 \\
    Town  & 0.017 & 0.330 &       & -0.510 & 0.263 &      & 0.017 & 0.321 &       & -0.610 & 0.274&    & 0.121 & 0.312 &       & -0.643 & 0.274 \\
    Village & 0.247 & 0.253 &       & -0.180 & 0.200 &       & 0.247 & 0.243 &       & -0.387 & 0.215&   & 0.383 & 0.235 &       & -0.434 & 0.215 \\
    AlarmCredit05  & 0.328 & 0.216 &       & 0.060 & 0.201 &      & 0.328 & 0.215 &       & 0.024 & 0.201 &     & 0.316 & 0.212 &       & 0.026 & 0.200 \\
    AlarmCredit10  & 0.316 & 0.205 &       & -0.121 & 0.177 &         & 0.316 & 0.203 &       & -0.201 & 0.181&   & 0.356 & 0.201 &       & -0.217 & 0.180 \\
    AlarmCredit15  & 0.227 & 0.136 &       & -0.115 & 0.121 &      & 0.227 & 0.135 &       & -0.123 & 0.124&    & 0.290 & 0.134 &       & -0.147 & 0.124 \\
    Deductible & -0.221 & 0.058 &       & 0.095 & 0.034 &       & -0.221 & 0.056 &       & 0.205 & 0.042&    & -0.322 & 0.064 &       & 0.235 & 0.044 \\
    Coverage & 0.782 & 0.054 &       & 0.048 & 0.037 &       & 0.782 & 0.053 &       & -0.010 & 0.041&    & 0.766 & 0.052 &       & -0.001 & 0.041 \\
    Spring &       &       &       & -0.110 & 0.106 &       &       &       &       & -0.064 & 0.104&    &       &       &       & -0.065 & 0.104  \\
    Summer &       &       &       & -0.040 & 0.099 &       &       &       &       & -0.023 & 0.097&    &       &       &       & -0.022 & 0.097 \\
    Fall  &       &       &       & 0.020 & 0.107 &        &       &       &       & 0.049 & 0.104&    &       &       &       & 0.053 & 0.104 \\
    Fire  &       &       &       & 0.533 & 0.085 &       &       &       &       & 0.468 & 0.085&    &       &       &       & 0.466 & 0.085 \\
    Water &       &       &       & 0.316 & 0.084 &      &       &       &       & 0.290 & 0.082&   &       &       &       & 0.288 & 0.082 \\
    ReportDelay &       &       &       & -0.001 & 0.001 &      &       &       &       & -0.001 & 0.001&    &       &       &       & -0.001 & 0.001 \\
    \hline
    \multicolumn{12}{l} {\textit{Zero-inflated Regression}}\\
    Intercept & -7.834 & 1.406 &       &       &       &     & -7.834 & 1.476 &       &       &  &     & -8.583 & 2.126 &       &       &  \\
    Deductible & 1.097 & 0.185 &       &       &       &       & 1.097 & 0.195 &       &       &  &   & 1.126 & 0.266 &       &       &  \\
    Coverage & -0.538 & 0.177 &       &       &       &       & -0.538 & 0.173 &       &       &  &   & -0.583 & 0.229 &       &       &  \\
    \multicolumn{12}{l} {\textit{One-inflated Regression}}\\
    Intercept & -7.411 & 1.507 &       &       &       &       & -7.411 & 1.557 &       &       & &   & -7.084 & 1.829 &       &       &  \\
    Deductible & 0.664 & 0.217 &       &       &       &        & 0.664 & 0.224 &       &       & &   & 0.577 & 0.266 &       &       &  \\
    Coverage & 0.020 & 0.182 &       &       &       &       & 0.020 & 0.184 &       &       &  &   & 0.016 & 0.201 &       &       &  \\
    \hline
    $\rho$ &       &       &       &       &    &    &   -0.290  &  0.034    &        &  &       &       &  -0.303  &     0.033   &       &   & \\
    \hline\hline
    \end{tabular}%
  \label{tab:modelest}%
\end{table}%
\end{landscape}

\subsection{Implications on Insurance Operations}
The previous section shows the statistical significance of the dependence between frequency and severity in the collective risk model. This section focuses on the substantive significance of the frequency-severity dependence and demonstrates its impacts on the decision-making in some key insurance operations (\cite{Frees2015}).

The first operation that we consider is underwriting and ratemaking. They are two basic functions in insurance companies and are closely related to each other. The former deals with the selection of risks, and latter deals with the determination of the price for the risks accepted. To achieve the underwriting profit target, the central task in underwriting and ratemaking is to quantify the risks of potential customers, which provides the insurer a risk score of policyholders to facilitate portfolio selection. To compare performance of the independence and the copula models, we look to the policyholders in the validation data of 2011 and examine which method leads to a more profitable portfolio construction.

\begin{figure}[htp]
  \begin{center}
   \includegraphics[width=0.45\textwidth,angle=270]{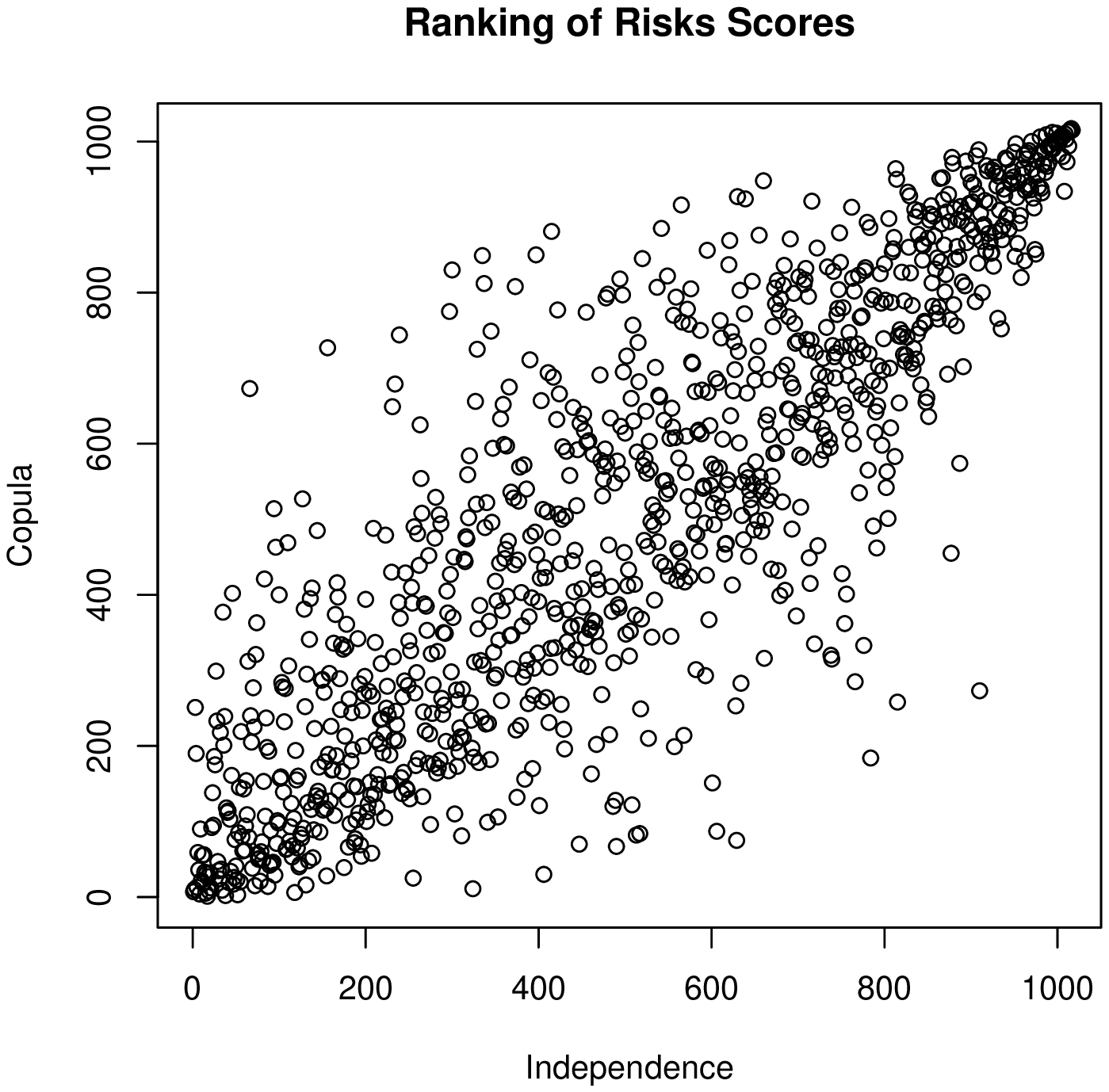}
   \includegraphics[width=0.45\textwidth,angle=270]{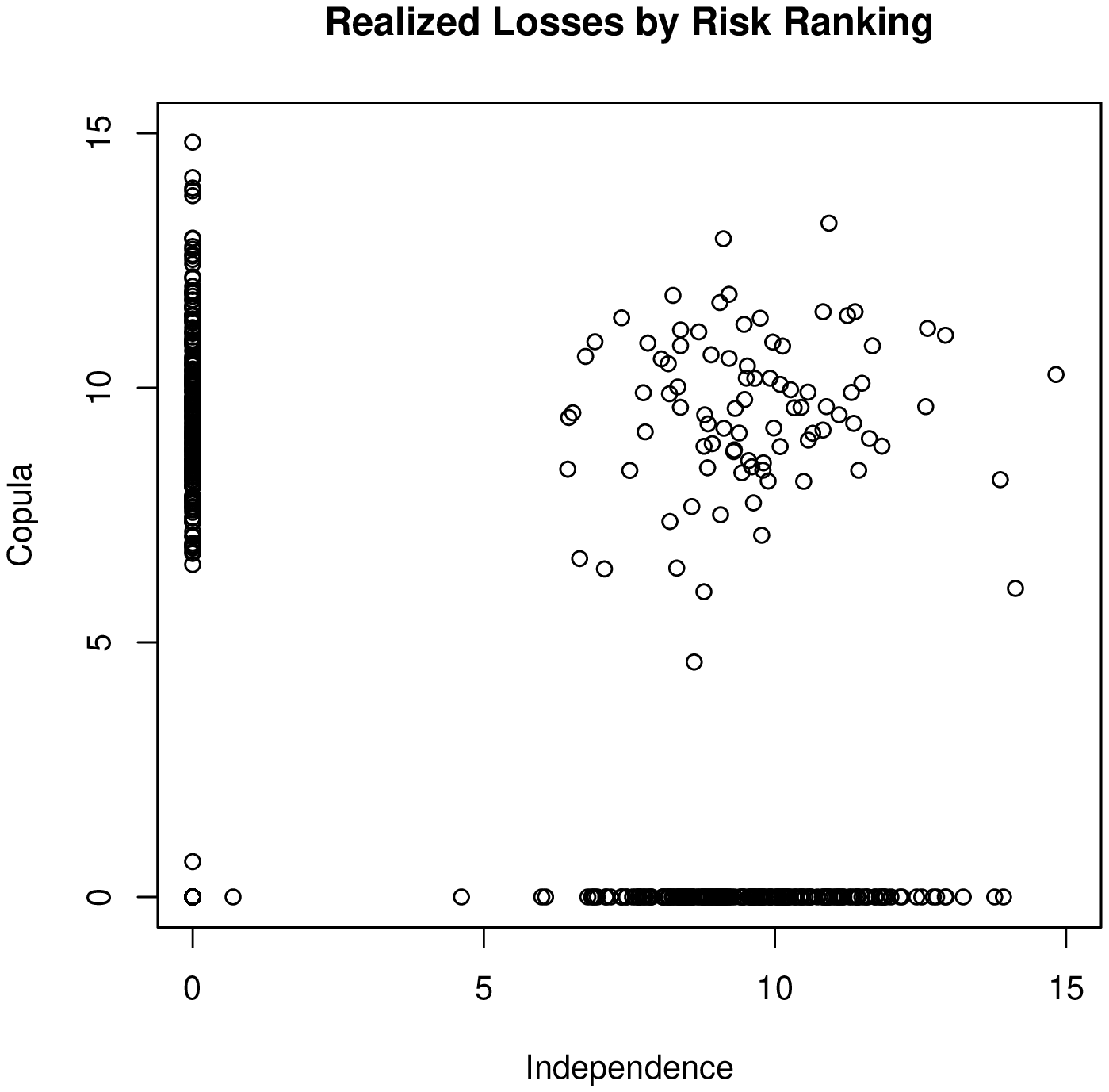}
   \includegraphics[width=0.45\textwidth,angle=270]{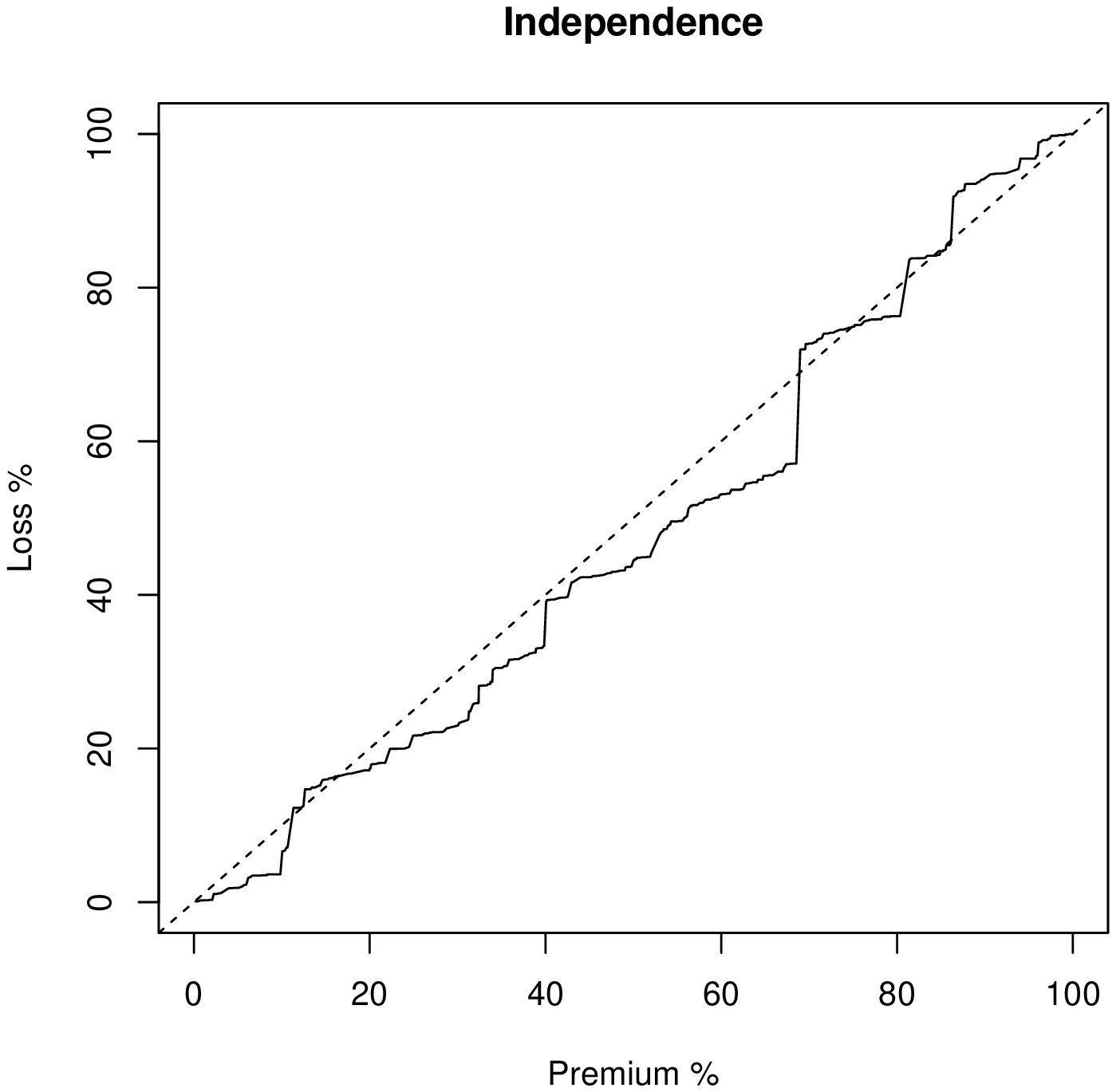}
   \includegraphics[width=0.45\textwidth,angle=270]{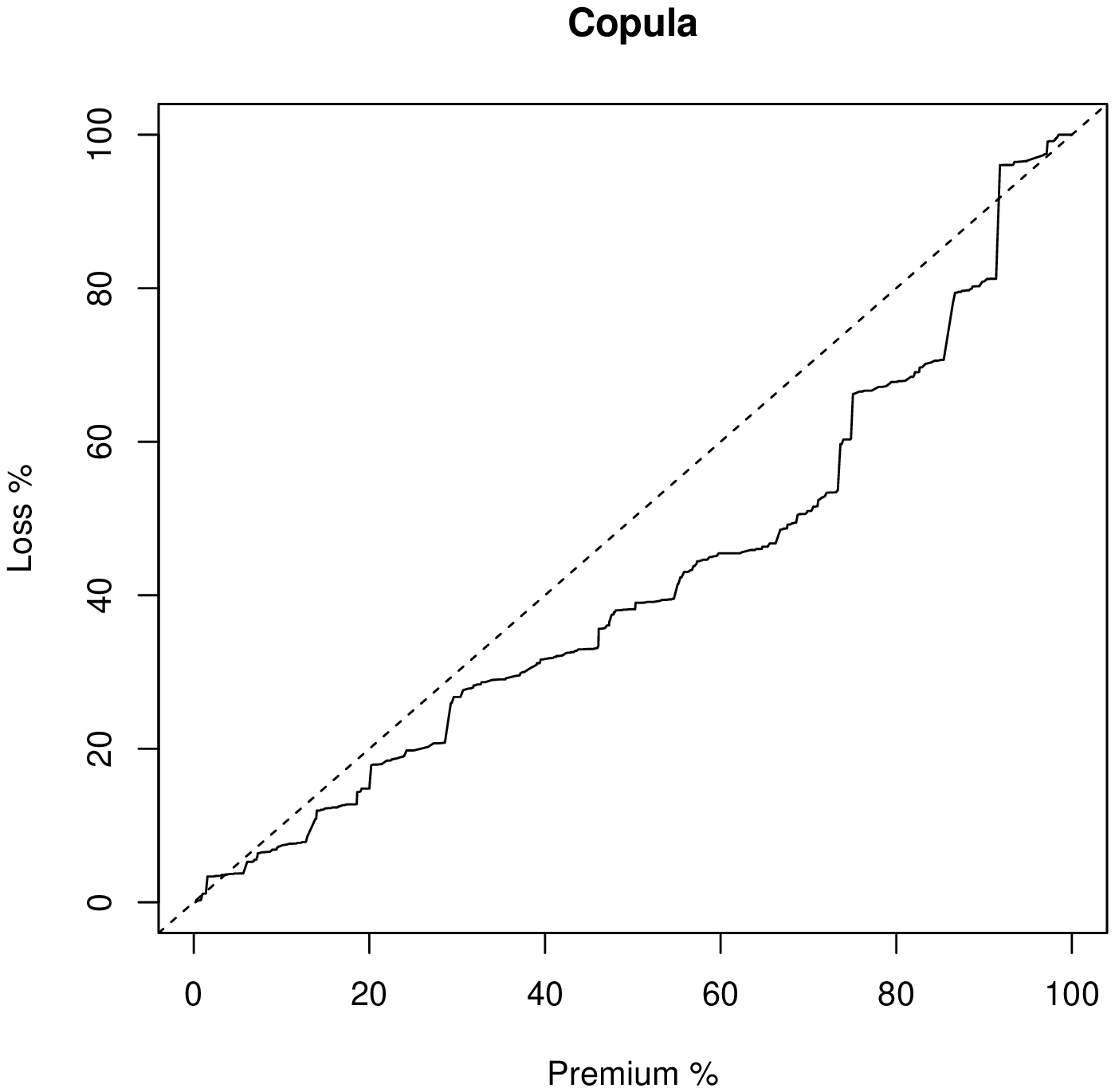}
  \end{center}
   \caption{{\color{black} Risk ranking and portfolio selection using the independent and the copula models. The top two figures compare risk score ranking and the corresponding realized losses between the independence and copula models respectively. The bottom two figures compare ordered Lorenz curves between the independence and copula models where the dashed line indicates perfect equality.}}
   \label{fig:Lorenz}
\end{figure}

For the purpose of underwriting, we use the coefficient of variation to measure the risk of policyholders. For each of the 1,017 policyholders in year 2011, we calculate the coefficient of variation of the loss cost, denoted $R_i=\sqrt{{\rm Var}[S_i]}/{\rm E}[S_i]$ for the $i$th policyholder. Given that the aggregate loss cost is specified using a collective risk model (\ref{equ:crm}), the mean and variance of $S$ is calculated by:
\begin{align*}
{\rm E}[S] &= {\rm E}[N{\rm E}[Y|N]]  \stackrel{\rm independence}=  {\rm E}[N]{\rm E}[Y] \\
{\rm Var}[S] &= {\rm E}[N{\rm Var}[Y|N]] + {\rm Var}[N{\rm E}[Y|N]] \stackrel{\rm independence}= {\rm E}[N]{\rm Var}[Y] + {\rm Var}[N] ({\rm E}[Y])^2
\end{align*}
The above calculation emphasizes the role of the dependency between the two building blocks, frequency and severity. We calculate the distribution of aggregate loss for each policyholder based on 10,000 Monte Carlo simulations. 
The upper panel of Figure \ref{fig:Lorenz} compares the risk ranking between the independence and the copula models. The first plot is the scatter plot of the ranking for each policyholder by the two methods. The second plot shows the realized aggregate losses (in log scale) with the same ranking from the two models. The risk scores from the two models are highly correlated yet there are considerable difference in their rankings.

To evaluate whether the risk ranking points to a profitable portfolio selection strategy, we display in the lower panel of Figure \ref{fig:Lorenz} the cumulative loss distribution ($F_L(R_i)$) versus the cumulative premium distribution ($F_P(R_i)$), both ordered by the riskiness of the policyholders $R_i$. This curve is known as the ordered Lorenz curve in \cite{FreesMeyersCummings2011}. In Figure \ref{fig:Lorenz}, the loss and premium distributions are calibrated using the realized losses of the policyholders and the actual premiums charged by the insurer in year 2011, respectively. The area between the curve and the 45 degree line is interpreted as an average profit or loss for the portfolio, with a convex curve for profit and a concave curve for loss. If one thinks of each underwriting strategy as retaining policies with riskiness less than or equal to $R_i$, the area represents an average profit in the sense that we are taking an expectation over all decision-making strategies. Furthermore, twice the area is known as the Gini index which thus has a natural economic interpretation. The Lorenz curve for the independence model is close to the 45 degree line. In contrast, the Lorenz curve for the copula model suggests a much higher average profit. Specifically, the Gini indices are 10.55\% and 33.24\% for the independence and the copula models, respectively. Therefore, a better underwriting strategy could be formed using the copula model, given that each policyholder is charged the contract premium.

We next compare the rates suggested by the independence and the copula models. A fair rate commensurate with the policyholder's risk mitigates adverse selection against the insurer. We perform a out-of-sample validation based on the Gini correlation in \cite{FreesMeyersCummings2011}. Two base premiums are considered, the constant premium and the contract premium. The former charges average cost to each policyholder, and the latter is the premium that the property fund charges based on the basic rating variables. Table \ref{tab:gini} presents the Gini correlation coefficients for the independence and the copula models. For both premium bases, the copula model shows a higher index, implying a more refined risk classification than the independence model.

\begin{table}[htbp]
  \centering
  \caption{Gini indices for independence and copula models$^\dag$}
    \begin{tabular}{lcc}
    \hline\hline
          & Independence & Copula \\
    \hline
    Constant Premium & 57.61 (6.57) & 63.24 (6.82) \\
    Contract Premium & 15.93 (8.81) & 26.27 (11.15)\\
    \hline\hline
    \multicolumn{3}{l}{$^\dag$ Standard errors are reported in parentheses.}
    \end{tabular}%
  \label{tab:gini}%
\end{table}%

The proposed copula model can also provide insights for the practice of claims reserving. In property casualty insurance, it is typical that a loss event won't be reported to the insurer immediately upon occurrence. For instance, a hail damage to the roof might be discovered by the policyholder several month later. After being reported, it further takes time for the insurer to decide coverage and finally settle the claim. Because of the long reporting and settlement delays, an insurer could be responsible for future payments associated with the loss events occurred in the policy period even post the expiration of the contract. Claims reserving or loss reserving is the process of estimating outstanding payments or the ultimate payments that an insurer is responsible for. Reserves are determined at both claim level and portfolio level (see, for example, \cite{AntonioBeirlant2008} and \cite{PigeonAntonioDenuit2014}). At claim level, an insurer estimates the amount for which a particular claim will ultimately be settled or adjudicated, also known as case reserve. At portfolio level, an insurer also estimates its future liabilities for the entire book of business. To emphasize its importance, loss reserves typically represent the largest liability item on the balance sheet of nonlife insurers.

\begin{figure}[htp]
  \begin{center}
   \includegraphics[width=0.45\textwidth,angle=270]{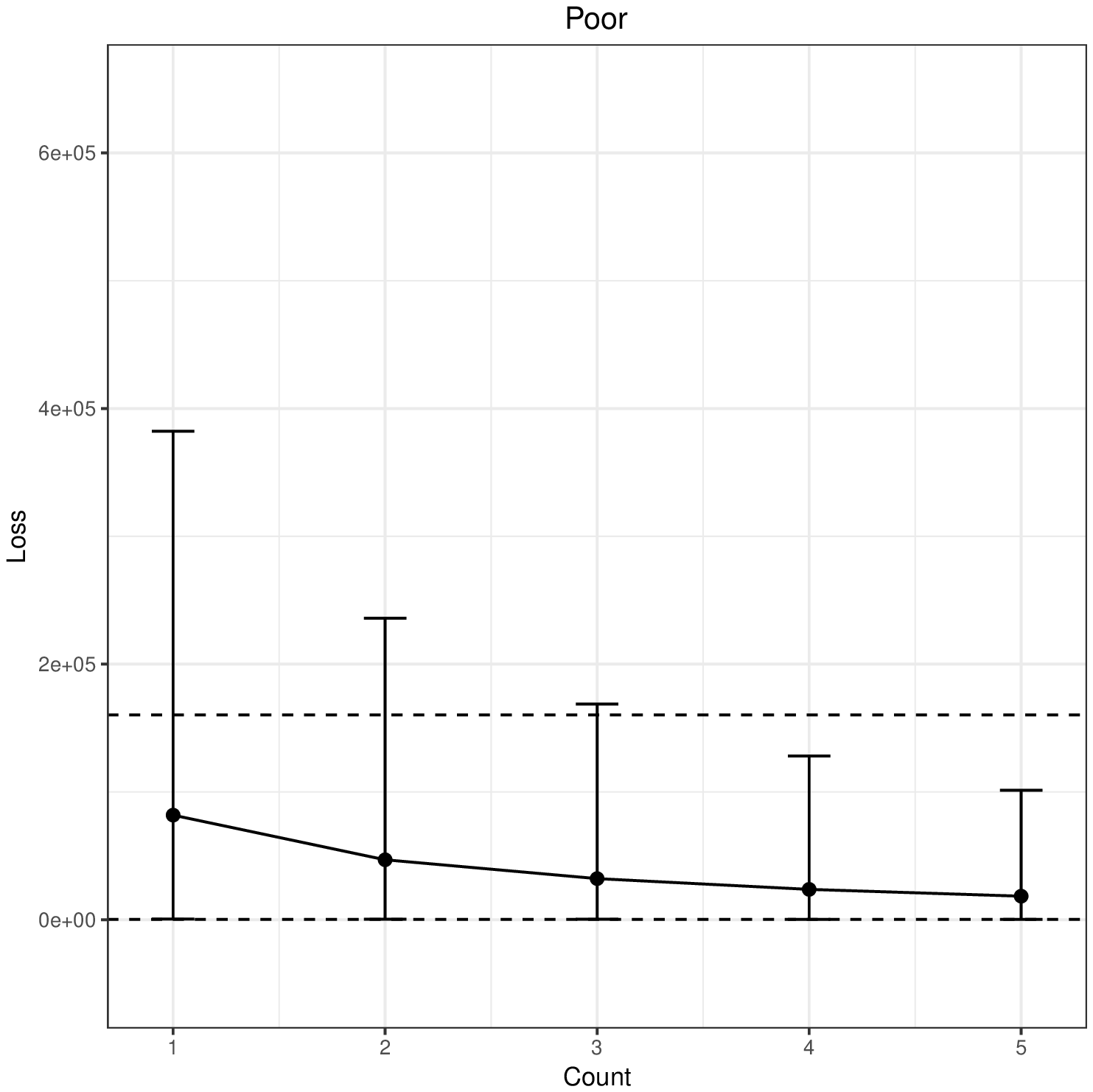}
   \includegraphics[width=0.45\textwidth,angle=270]{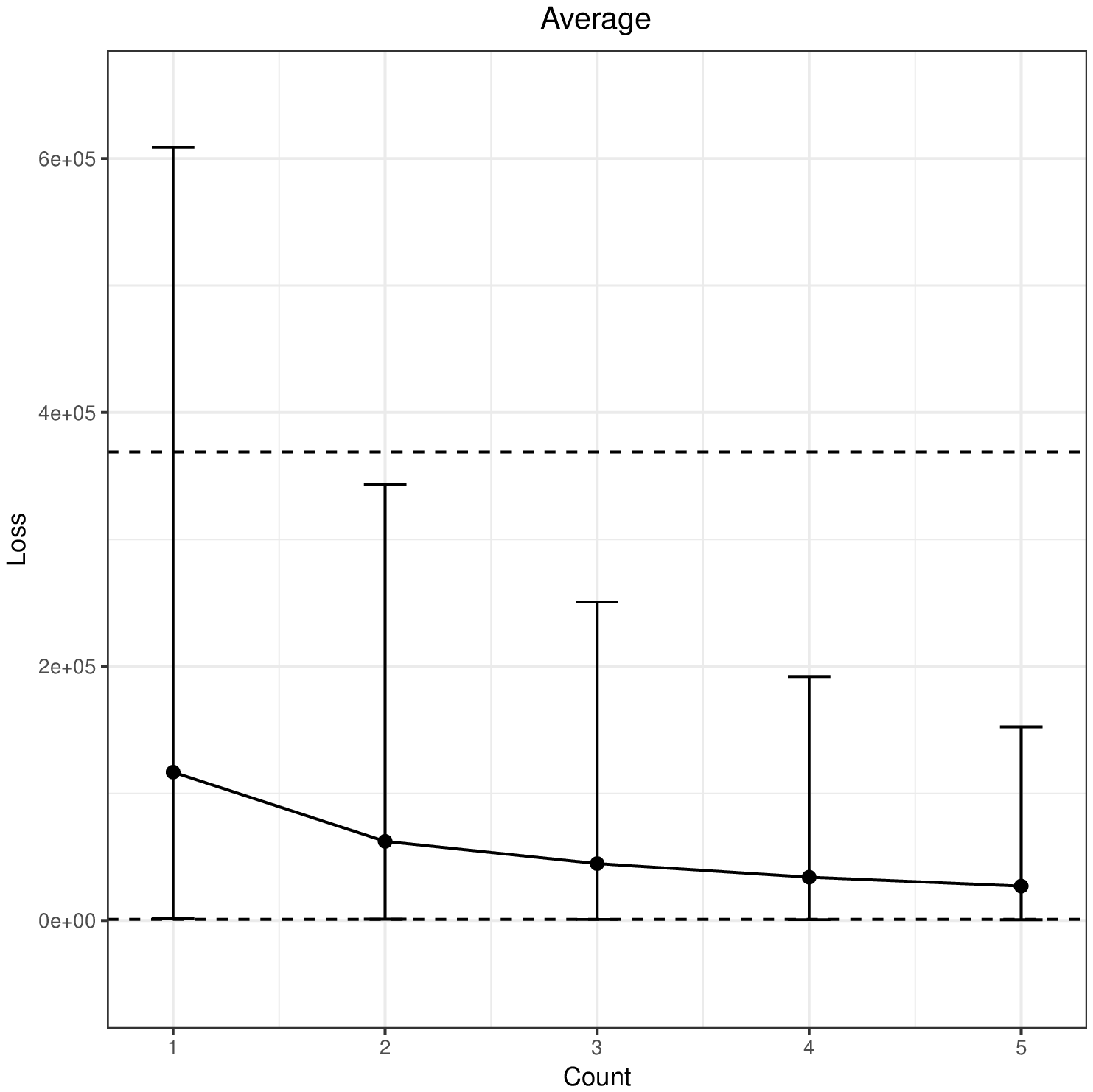}
   \includegraphics[width=0.45\textwidth,angle=270]{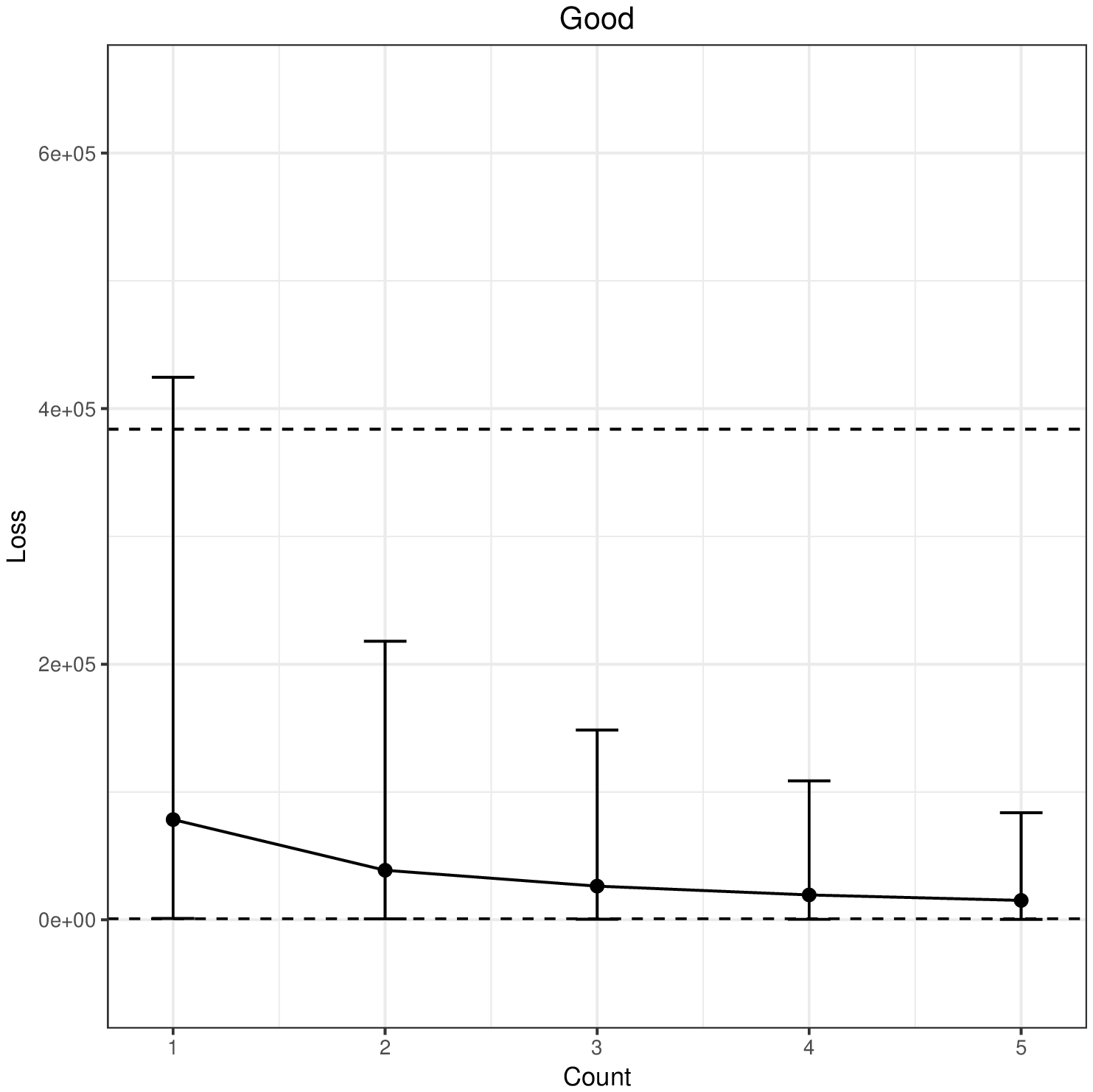}
   \includegraphics[width=0.45\textwidth,angle=270]{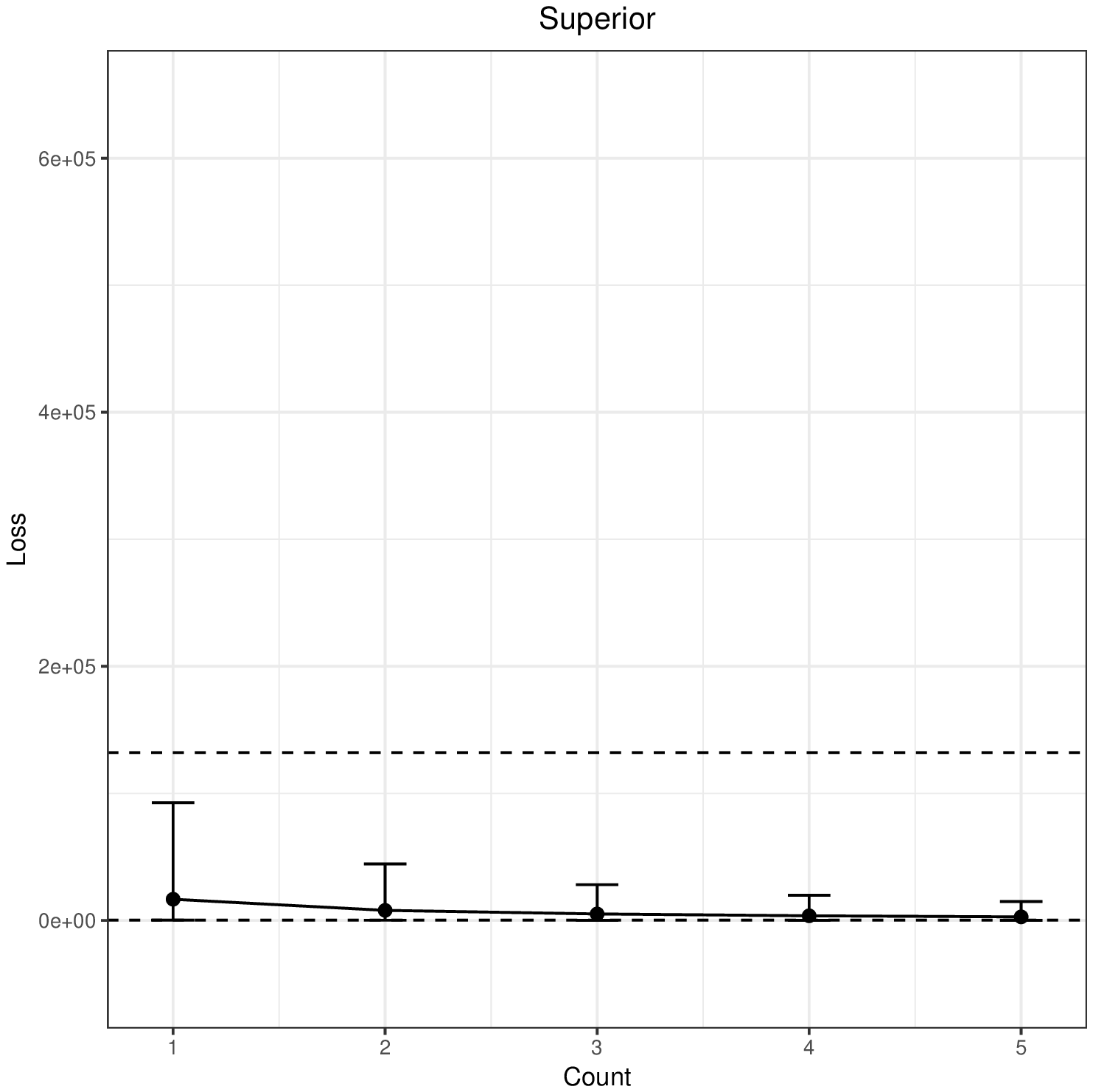}
  \end{center}
   \caption{Prediction interval of conditional claims amount for four representative risks.}
   \label{fig:respred}
\end{figure}

For reserving purposes, one is interested in the claims amount given occurrence of the loss events. As pointed out by \cite{WuthrichMerz2008}, because of the introduction of new supervisory guidelines (Solvency II) and financial reporting standards (IFRS 4 Phase II), the measurement of future cash flows and their uncertainty becomes more important. In this application, we examine the predictive distribution of $Y$ given $N$. For illustration, we display in Figure \ref{fig:respred} the 95\% prediction intervals of the claims amount for four representative risks, ``poor'', ``good'', ``average'', and ``superior''. The bar is determined by the 2.5th and 97.5th percentiles of the predictive distribution, and the solid dot indicates the predictive mean. The four risks are selected from the validation data based on the expected number of claims ${\rm E}(N)$. Specifically, they expect to have 2.37, 0.76, 0.37, 0.15 claims per year which corresponding to the 95th, 75th, 50th, 25th percentiles of the frequency distribution, respectively. For comparison, we impose the corresponding prediction interval from the independence model in the figure as indicated by the dashed line. First, as expected, the predictive distribution of claim amounts given frequency is skewed and long-tailed. This observation emphasizes that a range estimate of reserves is more informative than a point estimate for managers to set appropriate reserves, because an insurer doesn't want to overestimate nor underestimate its outstanding liabilities. Over-reserving could inflate the price and make the product less competitive, while under-reserving increases the solvency risk. Second, because of the significant negative relation between claim frequency and severity, the claims amount becomes smaller as the number of claims increases. A dynamic viewpoint is that an insurer updates its knowledge on the severity distribution based on frequency information. Third, it is apparent that ignoring the frequency-severity dependence will introduce significant bias in the reserving estimates. Under the independence assumption, not only that the claim severity is invariant with respect to claim frequency, but also the magnitude of the prediction could lead to poor decision making. For example, the results suggest that managers relying on the independence model tend to over reserve for better risks. In particular, the over-reserving risk is substantial for superior risks. As described earlier, there will be negative effects on both pricing and reserving. Over prediction of unpaid losses lead to increase in price which could cause the insurer to lose profitable business.

We further test the prediction of ultimate losses given occurrence for all the policyholders in the hold-out sample. To compare the prediction from the independence model to the copula model, we employ the continuous ranked probability score (CRPS) in \cite{GneitingRaftery2007} and \cite{Czado2009}. The CRPS is a proper scoring rule that assesses the quality of probabilistic forecasts. For reserving purpose, we focus on policyholders with at least one claims, and we evaluate the prediction of the aggregate loss distribution $f_{S|S>0}(s)$. The predictive distribution is derived for each policyholder based on 10,000 Monte Carlo simulations where the aggregated loss is generated conditional on occurrence of claims. Then the CRPS assigns a numerical score that measures the distance between the cumulative predictive distribution and the realized losses in the hold-out sample. For 73.34\% of the policies in the hold-out sample, the copula model outperforms the independence model. A binomial test suggest the superior prediction of the copula model to the independence model is statistically significant.

In the third application, we briefly demonstrate implications of the frequency-severity dependence on capital management. Insurance is a highly regulated industry. To mitigate solvency risk and protect public interest, insurers are required to hold minimum amount of risk capital as a buffer in case of some unexpected catastrophic events. We have already seen the consequences when the dependence between frequency and severity is unaccounted for at the individual policy level. This example emphasizes its relevance at the portfolio level since the risk capital is determined for the entire book of business.

To calculate the risk capital, we consider the value-at-risk (VaR), a risk measure widely used in the insurance and banking industry. The VaR focuses on the tail of the distribution, and specifically VaR$(\alpha)$ is defined as the $100\alpha$th percentile. Our interest is the aggregate losses for the insurance portfolio, defined as $L=\sum\nolimits_{i=1}^{m} S_i$, where $S_i$, the loss cost for policyholder $i$, is specified using the collective risk model (\ref{equ:crm}). The distribution of $L$ is estimated using 10,000 Monte Carlo simulations. Table \ref{tab:var} reports the risk measure at 90\%, 95\%, and 99\% levels for both the independence and copula models. To quantify the simulation uncertainty, we replicate the simulation 100 times to obtain the 95\% confidence interval. The results implies that ignoring the frequency-severity dependence in the collective risk model leads to significant underestimate of the tail risk for the portfolio. 


\begin{table}[htbp]
  \centering
  \caption{Value-at-Risk for the insurance portfolio (\$1,000)}
    \begin{tabular}{lccc}
    \hline\hline
    & 0.90 & 0.95 & 0.99 \\
    \cline{2-4}
    Independence & 39,556 & 69,124 & 314,854 \\
          & (38,961, 40,162) & (67,834, 70,521) & (300,348, 328,009)  \\
    Copula & 41,665 & 75,114 & 374,234 \\
          & (41,106,42,210) & (73,921,76,284) & (349,748, 397,509) \\
    Difference & 5.33\%  & 8.67\% &  18.86\%  \\
    \hline\hline
    \end{tabular}%
  \label{tab:var}%
\end{table}%


\section{Conclusion}

The two-part regression model based on compound distributions is commonly used in various disciplines, including insurance, economics, marketing, and  psychology, among others. The current practice is to perform a marginal regression on the primary (frequency) outcome, and a separate regression on the positive portion of the secondary (severity) outcome. This practice relies on the (conditional) independence assumption and causes significant biases in inference in the presence of frequency-severity dependence.

Motivated by the wide application of this type of model, this article represents the first attempt at accommodating the association between the frequency and severity components in the compound distribution and the associated regression models. We proposed the novel idea of using a parametric copula to construct the joint distribution of $N$ and $Y$ in the compound distribution. The copula regression is simple yet enjoys several advantages: First, the copula model allows for an arbitrary dependence between frequency and severity, and thus includes the (conditional) independence model as a special case. Second, separating the marginal from the joint distribution, the copula model can easily accommodate nonstandard marginal regressions for complicated data structure, for instance, regressions for zero/one-inflated data or the incomplete data due to censoring and truncation. Third, the parametric nature of the model implies straightforward likelihood-based inference and thus facilitates data-driven model specification and diagnostics, which is critical to the applications with complex and big data.

This work was motivated by the applications in insurance, where the complex and unique features of claims data provide a general setting to investigate the frequency-severity dependence in the context of the two-part model. For example, the standard count regression is not sufficient to capture the features in claim frequency; and the modifications on insurance coverage often cause observations to be incomplete. Although our empirical analysis emphasized the consequences of ignoring the frequency-severity dependence on the operations in insurance companies, the proposed model is general enough and ready to apply to other disciplines. It'll be interesting to see the implications of the frequency-severity dependence on decision making in other fields as well.

Finally, we conclude the paper with some discussions on the dependence between the frequency and severity in the proposed copula model. First, the proposed copula model relies on a simplifying assumption for the dependence, i.e. the association parameter in the copula is constant and does not vary across covariates. A potential extension is to use a conditional copula approach to allow the association in the copula to be dependent on covariates. See, for example, \cite{Patton2006modelling}, \cite{Acar2011dependence}, \cite{Veraverbeke2011}, \cite{fermanian2012time}, and \cite{castro2018time} for some recent development. {\color{black} We note that some domain knowledge is usually needed to support the conditional copula approach, for instance, the dependence among stock markets could be time-varying.} We leave it as a future research topic to investigate the conditional dependence in insurance data. \textcolor{black}{Second, we attribute the observed dependence in frequency and severity to unobserved heterogeneity. Regarding whether such relation is positive or negative, we think of this more as an empirical question to investigate. Often there are competing theories to support both positive and negative relationships. For the property insurance in our paper, one example of unobserved heterogeneity that induces dependence is weather related hazard. One can think of a geographical region that has frequent but modest storms versus another region that has infrequent but very severe storms. Another example of unobserved heterogeneity is the social-economic factors. One can think of some areas with frequent but minor crimes versus other areas with infrequent but severe crimes. Thus it is important for the model to offer the flexibility to accommodate both positive and negative relationship, and thus to allow for an empirical test of alternative theories.}

\newpage
\section*{SUPPLEMENTARY MATERIAL}
\renewcommand{\theequation}{S.\arabic{equation}}
\renewcommand{\thetable}{S.\arabic{table}}
\renewcommand{\thefigure}{S.\arabic{figure}}
\setcounter{figure}{0}
\setcounter{table}{0}
\setcounter{equation}{0}

\subsection*{S.1 Special Cases of the Copula Models}
We show that several widely used two-part models can be viewed in the proposed copula framework. The first is the hurdle model in the health economics literature (see, for instance, \cite{Mullahy1986} and \cite{PohlmeierUlrich1995}). The hurdle model considers the count measure
of health care utilisation as a result of two different decision processes. The first part specifies the decision
to seek care by individuals, and the second decision concerns the quantity of health care consumed which is partly determined by physicians. In the copula model, define $N=I(Y^{*}>0)$ and $Y=Y^*$ for a latent count variable $Y^{*}$. The copula for $N$ and $Y$ is shown to be $C(u,v)=\min{(u,v)}$. In this case, the copula model (\ref{equ:coprisk}) becomes:
\begin{align}
f_{N,\bm{Y}}(n,y_1,\ldots,y_n) &= \left[f_N(0)\right]^{I(n=0)}\left[f_N(1)\times f_{Y|N}(y|1)\right]^{I(n=1)}   \nonumber \\
& = \underbrace{\vphantom{\frac{F_{Y^*}(0)}{1-F_{Y^*}(0)}} \left[F_{Y^*}(0)\right]^{1-n}\left[1-F_{Y^*}(0)\right]^{n}}_\text{hurdle} \times \underbrace{\left[\frac{f_{Y^*}(y)}{1-F_{Y^*}(0)}\right]^{n} }_\text{truncated}.
\end{align}
This gives the standard hurdle model where the hurdle component can be a logit or a probit regression, and the truncated component is usually a Poisson or a negative binomial model. Governed by two different sets of parameters, the two components are separately and independently estimated. The same framework has also been used to study the semi-continuous health care expenditures where a log-linear or generalized linear model is often employed for the truncated component (see \cite{Mullahy1998}).

The second special case is the selection model. Assuming $N$ is a binary outcome and is determined by a latent continuous variable $N^{*}$ through relation $N = I(N^{*}>0)$, one has:
\begin{align*}
f_N(0)={\rm Pr}(N=0) &= {\rm Pr}(N^{*}\leq 0) = F_{N^{*}}(0) \\
f_N(1)={\rm Pr}(N=1) &= {\rm Pr}(N^{*}> 0) = 1-F_{N^{*}}(0)
\end{align*}
Denote $C(\cdot)$ as the copula that uniquely defines the joint distribution of $N^{*}$ and $Y$, the bivariate density of $N^{*}$ and $Y$ can be expressed as:
\begin{align*}
f_{N^{*},Y}(n^{*},y) = f_{N^{*}}(n^{*})f_Y(y)c(F_{N^{*}}(n^{*}),F_Y(y))
\end{align*}
where $c(u,v)=\dfrac{\partial^2}{\partial u \partial v}C(u,v)$. Then the distribution of $N$ and $Y$ is:
\begin{align*}
f_{N,Y}(0,y) &= \frac{\partial}{\partial y}{\rm Pr}(N^{*}\leq 0, Y\leq y) = \int_{-\infty}^{0} f_{N^{*},Y} (s,y) ds \\
& = f_Y(y)h(F_{N^{*}}(0),F_Y(y))\\
f_{N,Y}(1,y) &= \frac{\partial}{\partial y}{\rm Pr}(N^{*}>0, Y\leq y) = \int_{0}^{+\infty} f_{N^{*},Y} (s,y) ds \\
& = f_Y(y) (1-h(F_{N^{*}}(0),F_Y(y)))
\end{align*}
Note that the above is a selection model in that $Y$ is observed only if $N^{*}>0$. See \cite{Smith2003} and \cite{Prieger2002} for discussions on copula-based selection model. When $N^{*}$ and $Y$ are joint normal distribution, the copula model further reduces to the classic Heckman model (\cite{Heckman1979}).

Under this setting, the joint distribution of $(N,Y_1,\ldots,Y_N)$ is shown:
\begin{align*}
f_{N,\bm{Y}}(n,y_1,\ldots,y_n) &= [f_N(0)]^{1-n}[f_{N,Y}(1,y)]^n \\
& = F_{N^{*}}(0) ^{1-n} (1-F_{N^{*}}(0))^n \times \left(f_Y(y) \cfrac{1-h(F_{N^{*}}(0),F_Y(y))}{1-F_{N^{*}}(0)}\right)^n\\
& = \underbrace{ \vphantom{\cfrac{f_Y(y)}{f_N(1)}} f_{N}(0) ^{1-n} f_N(1) ^ n}_\text{\normalfont $f_N(n)$} \times \underbrace{\left(\cfrac{f_Y(y)}{f_N(1)} [1-h(F_{N}(0),F_Y(y))]\right)^n}_\text{\normalfont $f_{Y|N}(y|1)$}
\end{align*}
The model has a natural two-part interpretation where the joint distribution decouples into the product of frequency and severity distributions. However, the two components cannot be estimated separately because they are not independent with each other.

The last related case is the frequency-severity model in the actuarial science literature (\cite{Frees2014}), where the joint distribution of $(N,Y_1,\ldots,Y_N)$ is expressed as:
\begin{align} \label{equ:freqsev}
f_{N,\bm{Y}}(n,y_1,\ldots,y_n) = f_N(n) \times \prod_{j=1}^{n} f_{Y|N>0}(y_j).
\end{align}
The frequency component $f_N$ describes the number of claims and is specified as a count regression. The severity component $f_{Y|N>0}$ governs the size of claims given occurrence and employs generalized linear models to account for the skewness and heavy tails. The model assumes that given $N=n>0$, the distribution of $Y$ does not depend on $n$. Thus the two pieces can be estimated separately. Note the difference between distributions $f_{Y|N>0}$ and $f_{Y|N}$. To be more specific,
\begin{align*}
f_{Y|N>0}(y) = \frac{f_Y(y)}{1-F_N(0)}, ~~~~{\rm and}~~~~ f_{Y|N}(y|n) = \frac{f_{N,Y}(n,y)}{f_N(n)}.
\end{align*}

The frequency-severity model can also be formulated using the proposed copula framework. Consider the case where $N$ and $Y$ is only related through the relation between $D$, which is defined as $D = I(N>0)$, and $Y$. That is, $N$ and $Y$ are independent conditional on $D$. Let $C(\cdot)$ denote the copula that defines the joint distribution of $D$ and $Y$, we have
\begin{align*}
f_{D,Y}(d,y) = \left\{
\begin{array}{ll}
f_Y(y)h(F_{D}(0),F_Y(y)), & d=0 \\
f_Y(y)[1-h(F_{D}(0),F_Y(y))], & d=1 \\
\end{array}
\right..
\end{align*}
Then the joint distribution of $N$ and $Y$ can be shown as:
\begin{align} \label{equ:freqsev2}
f_{N,Y}(n,y) &= {\rm E}_{D}\left[f_{N,Y|D}(n,y|d)\right] \nonumber\\
& =\sum_{d=0}^{1} f_{N|D}(n|d)f_{Y|D}(y|d) f_{D}(d)  \nonumber\\
&=[f_Y(y)h(F_N(0),F_Y(y))]^{I(n=0)} [f_N(n)f_{Y|N>0}(y|N>0)]^{I(n>0)}
\end{align}
Using (\ref{equ:freqsev2}), the copula model (\ref{equ:coprisk}) becomes the standard frequency-severity model (\ref{equ:freqsev}), which further reduces to (\ref{equ:indrisk}) when $D$ and $Y$ are independent. It is important to note that models (\ref{equ:indrisk}) and (\ref{equ:freqsev}) are indistinguishable because $Y$ is not completely observed and thus only the partial information corresponding to $N>0$ is available for inference.

{\color{black}

\subsection*{S.2 Additional Simulation Studies}

In this section, we conduct additional simulation studies to examine the estimation performance for censored and truncated data as described in Section {\ref{sec:incomplete}}.} We set the deductible $d=50$ and policy limit $l=10,000$, which are roughly the 20th and 90th percentiles of the marginal distribution of $Y$. Estimation results using independence estimation and joint MLE are summarized in Table \ref{tab:estincom_censored} and Table \ref{tab:estincom_truncated} for the censored and truncated data respectively. In general, the joint MLE performs well for both cases of incomplete data. Higher estimation uncertainty is associated with the truncated data compared to the censored data due to less amount of information available. The comparison with independence estimation emphasizes the significant bias when ignoring the dependence between frequency and severity. Similar to the complete data, when data are censored, independence estimation only introduces persistent bias in the severity model while the frequency model can still be consistently estimated. However, large bias is anticipated and observed in both frequency and severity models when data are truncated, and the bias won't disappear when sample size increases.

\begin{table}[htbp]
	\centering
	\caption{Estimation results for censored data using independence estimation and joint MLE}
	\begin{tabular}{lrrrrrrr}
		\hline\hline
		& \multicolumn{3}{c}{Independence}&& \multicolumn{3}{c}{Joint MLE}  \\
		\cline{2-4}\cline{6-8}
		Low Dependence    & Mean & Relative Bias & RMSE &   & Mean & Relative Bias & RMSE     \\
		\hline
		$\beta_0^f$ =-1.5  & -1.508 & 0.005 & 0.116 & & -1.504 & 0.003 & 0.111    \\
		$\beta_1^f$ = 2.5  & 2.513 & 0.005 & 0.144 & & 2.507 & 0.003 & 0.140    \\
		$\beta_2^f$ = 1    & 0.998 & -0.002 & 0.073 & & 0.997 & -0.003 & 0.071   \\
		$\beta_0^s$ = 5    & 5.084 & 0.017 & 0.115 & & 4.997 & -0.001 & 0.093  \\
		$\beta_1^s$ = -2.5  & -2.561 & 0.025 & 0.139 & & -2.499 & -0.000 & 0.121   \\
		$\beta_2^s$= 5     & 4.998 & -0.000 & 0.077 & & 5.005 & 0.001 & 0.070   \\
		$\alpha$ = 2    & 2.064 & 0.032 & 0.127 & & 2.015 & 0.008 & 0.113   \\
		$\rho$ = 0.1   &&&& & 0.097 & -0.030 & 0.042   \\
		\hline
		Medium Dependence   & Mean & Relative Bias & RMSE &  & Mean & Relative Bias & RMSE    \\
		\hline
		$\beta_0^f$ =-1.5  & -1.508 & 0.005 & 0.116 & & -1.505 & 0.004 & 0.106   \\
		$\beta_1^f$ = 2.5  & 2.513 & 0.005 & 0.144 & & 2.507 & 0.003 & 0.131   \\
		$\beta_2^f$ = 1    & 0.998 & -0.002 & 0.073 & & 0.998 & -0.002 & 0.070   \\
		$\beta_0^s$ = 5    & 5.421 & 0.084 & 0.432  & & 4.996 & -0.001 & 0.089  \\
		$\beta_1^s$ = -2.5 & -2.762 & 0.105 & 0.300 & & -2.495 & -0.002 & 0.128   \\
		$\beta_2^s$= 5     & 4.940 & -0.012 & 0.097 & & 5.002 & 0.000 & 0.064  \\
		$\alpha$ = 2      & 2.370 & 0.185 & 0.393 & & 2.019 & 0.010 & 0.127 \\
		$\rho$ = 0.5   &&&& & 0.499 & -0.002 & 0.030 \\
		\hline
		High Dependence   & Mean & Relative Bias & RMSE &   & Mean & Relative Bias & RMSE    \\
		\hline
		$\beta_0^f$ =-1.5 & -1.508 & 0.005 & 0.116 & & -1.506 & 0.004 & 0.083  \\
		$\beta_1^f$ = 2.5 & 2.513 & 0.005 & 0.144 & & 2.508 & 0.003 & 0.097   \\
		$\beta_2^f$ = 1   & 0.998 & -0.002 & 0.073 & & 1.000 & 0.000 & 0.066  \\
		$\beta_0^s$ = 5   & 5.727 & 0.145 & 0.737 & & 4.995 & -0.001 & 0.070   \\
		$\beta_1^s$ = -2.5 & -2.952 & 0.181 & 0.492 & &-2.491 & -0.004 & 0.117   \\
		$\beta_2^s$= 5    & 4.894 & -0.021 & 0.153 & & 4.999 & -0.000 & 0.056   \\
		$\alpha$ = 2   & 2.945 & 0.473 & 0.985 & & 2.013 & 0.007 & 0.095  \\
		$\rho$ = 0.9   &&&&& 0.900 & 0.000 & 0.006  \\
		\hline\hline
	\end{tabular}%
	\label{tab:estincom_censored}%
\end{table}%

\begin{table}[htbp]
	\centering
	\caption{Estimation results for truncated data using independence estimation and joint MLE}
	\begin{tabular}{lrrrrrrr}
		\hline\hline
		& \multicolumn{3}{c}{Independence}&& \multicolumn{3}{c}{Joint MLE}  \\
		\cline{2-4}\cline{6-8}
		Low Dependence    & Mean & Relative Bias & RMSE &   & Mean & Relative Bias & RMSE     \\
		\hline
		$\beta_0^f$ =-1.5  & -1.512 & 0.008 & 0.190  && -1.508 & 0.005 & 0.183  \\
		$\beta_1^f$ = 2.5  & 2.507 & 0.003 & 0.156  && 2.494 & -0.002 & 0.151  \\
		$\beta_2^f$ = 1    & 1.001 & 0.001 & 0.198 & &  0.997 & -0.003 & 0.191  \\
		$\beta_0^s$ = 5    & 5.085 & 0.017 & 0.137 & & 5.004 & 0.001 & 0.117  \\
		$\beta_1^s$ = -2.5  & -2.562 & 0.025 & 0.139 && -2.501 & 0.000 & 0.124  \\
		$\beta_2^s$= 5     & 4.997 & -0.001 & 0.097 && 4.998 & -0.000 & 0.090  \\
		$\alpha$ = 2    & 2.067 & 0.034 & 0.137 & & 2.019 & 0.010 & 0.118   \\
		$\rho$ = 0.1   & && & & 0.097 & -0.031 & 0.045  \\
		\hline
		Medium Dependence   & Mean & Relative Bias & RMSE &  & Mean & Relative Bias & RMSE    \\
		\hline
		$\beta_0^f$ =-1.5  & -1.440 & -0.040 & 0.172 & & -1.500 & -0.000 & 0.153  \\
		$\beta_1^f$ = 2.5  & 2.515 & 0.006 & 0.154  & & 2.500 & -0.000 & 0.137  \\
		$\beta_2^f$ = 1    & 0.929 & -0.071 & 0.185 & & 0.987 & -0.013 & 0.152  \\
		$\beta_0^s$ = 5    & 5.384 & 0.077 & 0.401 & & 4.997 & -0.001 & 0.096  \\
		$\beta_1^s$ = -2.5 & -2.753 & 0.101 & 0.291 & & -2.500 & -0.000 & 0.132  \\
		$\beta_2^s$= 5     & 4.971 & -0.006 & 0.097 & & 5.001 & 0.000 & 0.073  \\
		$\alpha$ = 2     & 2.345 & 0.172 & 0.371 & & 2.017 & 0.008 & 0.130  \\
		$\rho$ = 0.5   & && & & 0.499 & -0.003 & 0.033  \\
		\hline
		High Dependence   & Mean & Relative Bias & RMSE &   & Mean & Relative Bias & RMSE    \\
		\hline
		$\beta_0^f$ =-1.5  & -1.403 & -0.065 & 0.172 & & -1.508 & 0.005 & 0.115  \\
		$\beta_1^f$ = 2.5  & 2.521 & 0.008 & 0.148 & & 2.499 & -0.000 & 0.108  \\
		$\beta_2^f$ = 1     & 0.888 & -0.112 & 0.190 & &  1.002 & 0.002 & 0.114  \\
		$\beta_0^s$ = 5     & 5.671 & 0.134 & 0.685 & &  4.996 & -0.001 & 0.076  \\
		$\beta_1^s$ = -2.5 &-2.941 & 0.177 & 0.485 & & -2.499 & -0.000 & 0.135  \\
		$\beta_2^s$= 5    & 4.946 & -0.011 & 0.148 & & 5.001 & 0.000 & 0.068  \\
		$\alpha$ = 2     & 2.863 & 0.432 & 0.906 & & 2.013 & 0.007 & 0.098  \\
		$\rho$ = 0.9   &&& & & 0.900 & -0.001 & 0.007  \\
		\hline\hline
	\end{tabular}%
	\label{tab:estincom_truncated}%
\end{table}%

{\color{black}
\subsection*{S.3 Covariates, Diagnostic Analysis, and Robustness Check}  

Table \ref{tab:covaraites} summarizes the covariates (both variable description and descriptive statistics) in the regression analysis. The frequency model uses the policy-specific covariates only, the severity model uses both policy-specific and claim-specific covariates.

\begin{table}[htbp]
  \centering
  \caption{Description and summary statistics of covariates}
    \begin{tabular}{llrr}
    \hline\hline
    Variable & Description & Mean & SD \\
    \hline
    \textit{Policy-specific}  &       &       &  \\
    City  & Indicator of city & 0.144 &  \\
    County & Indicator of county & 0.059 &  \\
    School & Indicator of school & 0.292 &  \\
    Town  & Indicator of town & 0.167 &  \\
    Village & Indicator of village & 0.232 &  \\
    Misc  & Indicator of miscellaneous entities (example: fire station) & 0.106 &  \\
    AlarmCredit00  & Indicator of 0\% alarm credit & 0.346 &  \\
    AlarmCredit05  & Indicator of 5\% alarm credit & 0.064 &  \\
    AlarmCredit10  & Indicator of 10\% alarm credit & 0.076 &  \\
    AlarmCredit15  & Indicator of 15\% alarm credit & 0.513 &  \\
    Deductible & Log of deductible amount in dollars & 7.208 & 1.174 \\
    Coverage & Log of coverage amount in millions of dollars & 2.261 & 1.976 \\
    \hline
    \textit{Claim-specific} &       &       &  \\
    Spring & Indicator of whether occurrence in spring & 0.211 &  \\
    Summer & Indicator of whether occurrence in summer & 0.391 &  \\
    Fall  & Indicator of whether occurrence in fall & 0.208 &  \\
    Winter & Indicator of whether occurrence in winter & 0.190 &  \\
    Fire  & Indicator of fire peril of loss & 0.290 &  \\
    Water & Indicator of water peril of loss & 0.281 &  \\
    Other & Indicator of other perils of loss & 0.429 &  \\
    ReportDelay & Log of delay in accident report in days & 2.725 & 1.369 \\
    \hline\hline
    \end{tabular}%
  \label{tab:covaraites}%
\end{table}%

}

We follow the procedures described in Section \ref{subsec:inference} to perform diagnostic checking. For the frequency model, we investigate the popular Poisson and negative binomial regressions, along with their zero inflated and zero-one inflated versions. Table \ref{tab:chifreq} presents the goodness-of-fit statistics for the alternative count regressions. The poor fit of the Poisson model is not surprising due to its incapability to handle zero-inflation and over dispersion. The negative binomial regression and the zero-inflated models account for zero inflation but do not accommodate the probability mass at one. The Pearson's $\chi^2$ statistic is in favor of the zero-one inflated negative binomial model.

\begin{table}[htbp]
	\centering
	\caption{Goodness-of-fit statistics for alternative frequency models}
	\begin{tabular}{lrrrrrrr}
		\hline\hline
		& {Empirical} & {Poisson} & {NB} & {ZIP} & {ZINB} & {ZOIP} & {ZOINB} \\
		\hline
		\% of zeros & 0.681 & 0.643 & 0.688 & 0.678 & 0.689 & 0.679 & 0.683 \\
		\% of ones & 0.194 & 0.211 & 0.178 & 0.173 & 0.176 & 0.196 & 0.195 \\
		\hline
		$\chi^2$-stat &       & 111.73 & 38.67 & 97.33 & 41.09 & 40.41 & 32.13 \\
		\hline\hline
	\end{tabular}%
	\label{tab:chifreq}%
\end{table}%

\begin{figure}[htp]
	\begin{center}
		\includegraphics[width=0.4\textwidth,angle=270]{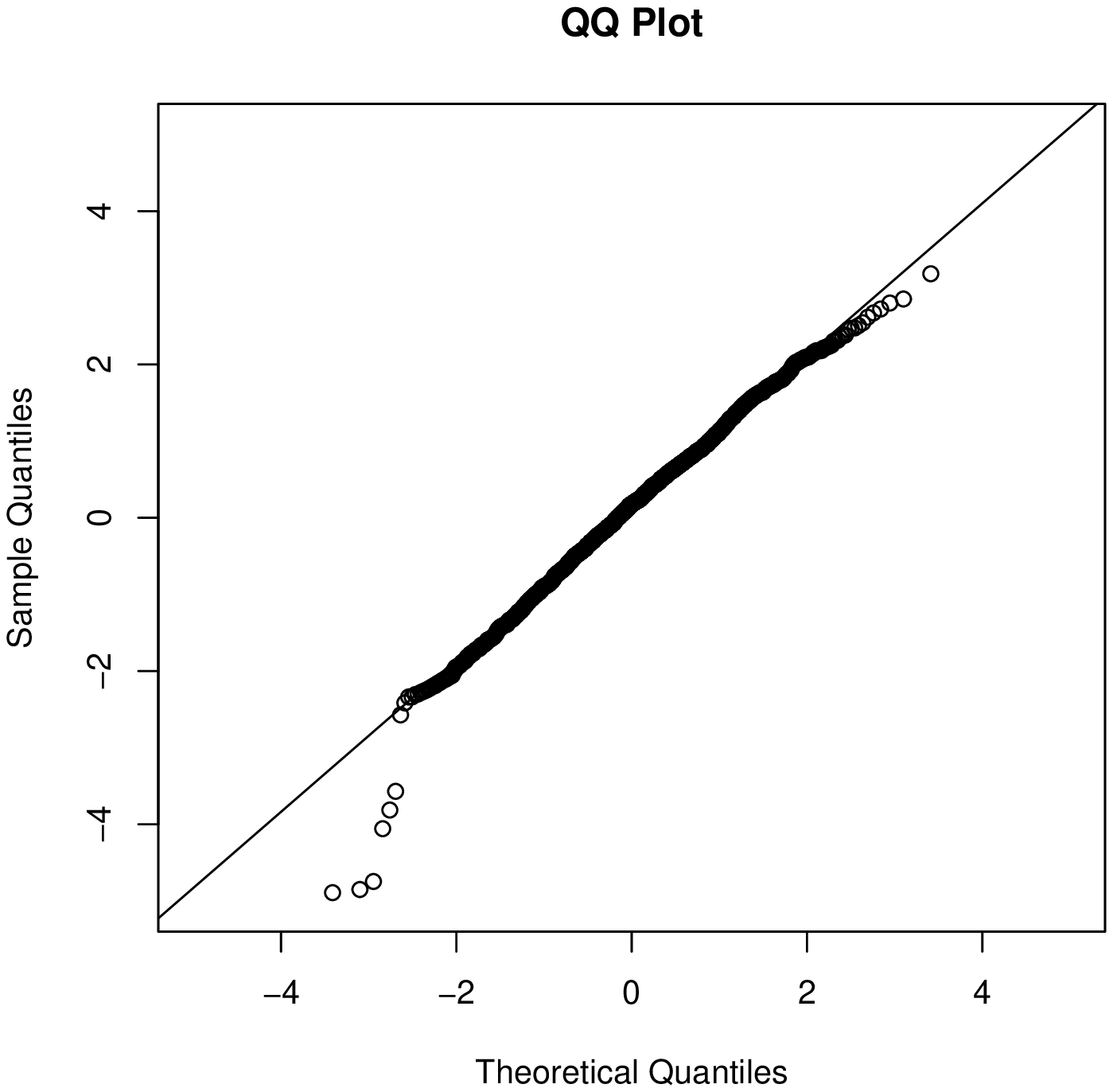}
		\includegraphics[width=0.4\textwidth,angle=270]{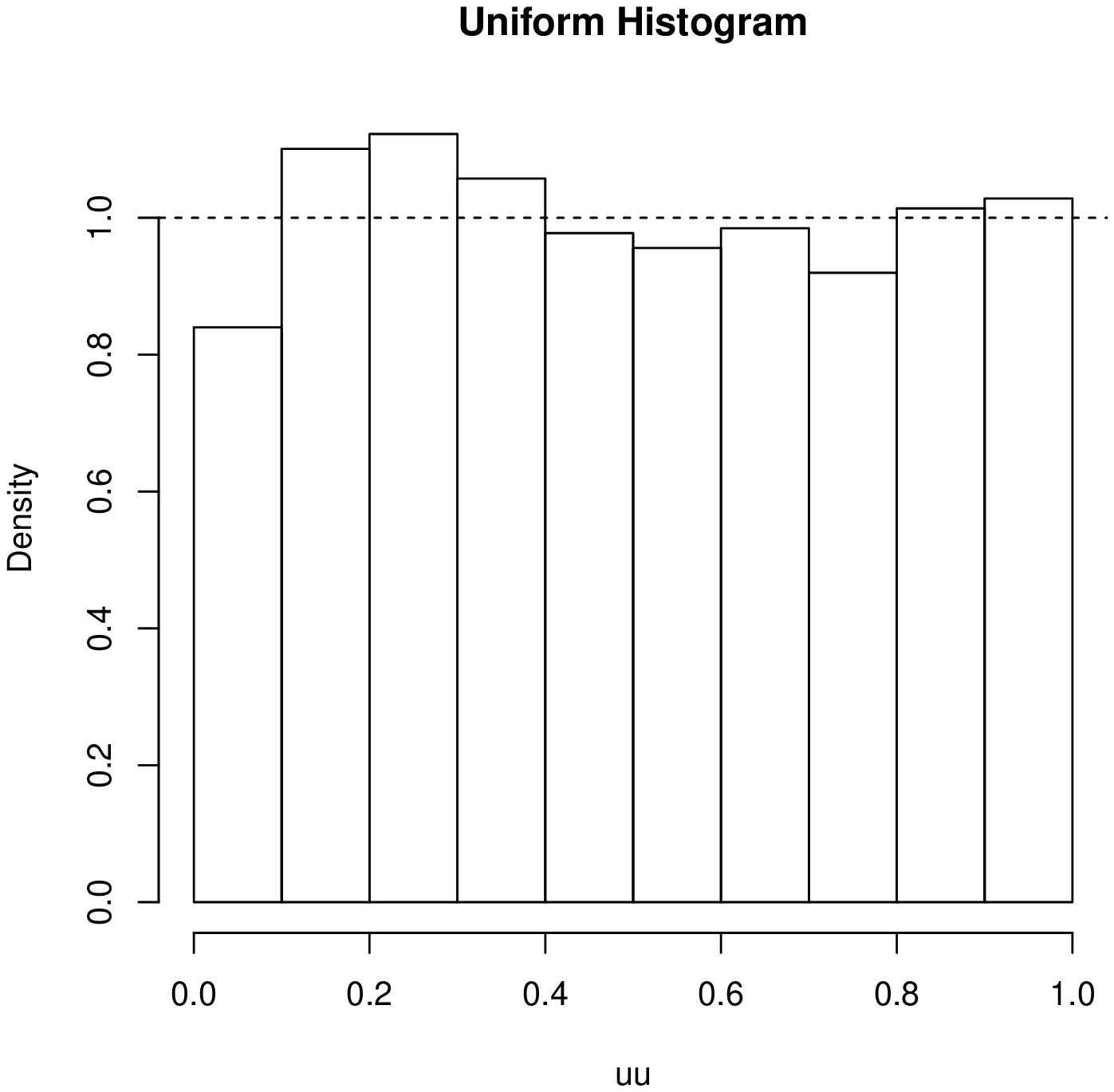}
	\end{center}
	\caption{Diagnostic checking for the conditional severity. The left panel shows the QQ plot for the normal scores. The right panel shows the histogram of the Cox-Snell residuals.}
	\label{fig:qqplot}
\end{figure}

As discussed in Section \ref{subsec:inference}, the marginal distribution of $Y$ and the copula cannot be separately identified. Therefore, the model checking is based on the conditional distribution $f_{Y|N}$. Figure \ref{fig:qqplot} displays the QQ plot of the normal scores $\Phi^{-1}\left(\widehat{F}_{Y|N}(y_{ij}|n_i)\right)$ and the histogram of the Cox-Snell residuals $\widehat{F}_{Y|N}(y_{ij}|n_i)$. There is no irregular pattern in the plots, suggesting that the severity distribution and the bivariate copula are well specified. The slight deviation in the left tail of the QQ plot is caused by five claims that have one dollar as claim amount, which bear minimal impact in insurance operation. A formal uniform test for the Cox-Snell residuals is provided in Table \ref{tab:unitest}, and the large $p$-values support the uniformity of the residuals.

\begin{table}[htbp]
	\centering
	\caption{Uniform test for the Cox-Snell residuals}
	\begin{tabular}{lrr}
		\hline\hline
		& \multicolumn{1}{l}{Statistics} & \multicolumn{1}{l}{$p$-value} \\
		\hline
		Kolmogorov-Smirnov & 0.020 & 0.613\\
		Cramer-von Mises & 0.085 & 0.665 \\
		Anderson-Darling & 0.943 & 0.388 \\
		\hline\hline
	\end{tabular}%
	\label{tab:unitest}%
\end{table}%

In the analysis of Wisconsin property fund data, we have found significant negative association between the number of claims and the size of claims for local government entities. One question of interest is how robust this negative relationship is with respect to the year of observations. Recall that we have used data of 2009 and 2010 to build the model and reserved the data of 2011 for hold-out sample validation. For the purpose of robustness check, we reestimate the proposed copula model with all three years of data. To focus on the frequency-severity dependence, we consider the same set of copula candidates as in Table \ref{tab:aic}. The results are summarized in Table \ref{tab:aicrobust}. First, the magnitude and the statistical significance of the negative association is in line with the estimates in Table \ref{tab:aic}. Second, the same favorite, Gaussian copula, is selected by the information criteria and goodness-of-fit statistic.

\begin{table}[htbp]
  \centering
  \caption{Robust check of goodness-of-fit statistics for various copula models}
    \begin{tabular}{lrrrrr}
    \hline\hline
          & {Kendall's $tau$} & {LogLik} & {AIC} & {BIC} & {Pearson's $\chi^2$} \\
    \hline
    Independence &       & -23,722 & 47,522 & 47,769 &  \\
    Gaussian & -0.20 & -23,667 & 47,412 & 47,659 & 110.20 \\
    $t$     & -0.20 & -23,667 & 47,412 & 47,659 & 110.20 \\
    Clayton90 & -0.08 & -23,678 & 47,434 & 47,682 & 87.82 \\
    Clayton270 & -0.35 & -23,689 & 47,456 & 47,703 & 66.18 \\
    Gumbel90 & -0.29 & -23,671 & 47,419 & 47,666 & 103.06 \\
    Gumbel270 & -0.08 & -23,692 & 47,462 & 47,709 & 60.24 \\
    Frank90/270 & -0.23 & -23,683 & 47,444 & 47,691 & 78.14 \\
    Joe90 & -0.37 & -23,692 & 47,461 & 47,708 & 61.06 \\
    Joe270 & -0.04 & -23,698 & 47,475 & 47,722 & 47.26 \\
    \hline\hline
    \end{tabular}%
  \label{tab:aicrobust}%
\end{table}%

{\small
\bibliographystyle{chicago}
\bibliography{ref_FreqSev}
}

\end{document}